\documentclass[preprint,amsmath,amssymb,nofootinbib]{revtex4}
\usepackage[dvips]{graphics}
\usepackage{graphics}
\usepackage{multirow}
\usepackage{amsmath}
\usepackage{array}
\usepackage{booktabs}
\makeatletter
\renewcommand{\fnum@figure}{Fig.~ \thefigure}
\makeatother
\usepackage{pstcol}
\usepackage{amsfonts}
\usepackage{bm}
\usepackage{amsmath}
\usepackage{amssymb}
\usepackage{color}
\usepackage[all]{xy}
\usepackage{comment}
\usepackage{epstopdf,epsfig}
\def\be{\begin{equation}}
\def\ee{\end{equation}}
\def\bea{\begin{eqnarray}}
\def\eea{\end{eqnarray}}
\usepackage{float}
\usepackage{xcolor}
\usepackage{xcolor}
\definecolor{fucsia}{HTML}{FF00FF}
\usepackage{graphicx}
\usepackage{makecell}
\usepackage{multirow}
\usepackage{subcaption}
\usepackage{setspace}
\usepackage{microtype}
\usepackage[
    colorlinks=true,
    linkcolor=blue,     
    citecolor=red,      
]{hyperref}

\usepackage{ragged2e}

\makeatletter
\def\@makecaption#1#2{%
  \vskip 8pt 
  \begingroup
    \small
    \justifying
    \textbf{#1.} #2\par
  \endgroup
}
\makeatother

\begin{document}

\title{Probing Black Hole Thermodynamics and Microstructure\\ via the Shadow of Sagittarius $A^{*}$}


\author{Jose Miguel Ladino$^1$, Carlos E. Romero-Figueroa$^1$, and Hernando Quevedo$^{1,2,3}$}
\email{\raggedright
miguel.ladino@correo.nucleares.unam.mx;
carlosed.romero@correo.nucleares.unam.mx;
quevedo@nucleares.unam.mx
}
\affiliation{\hspace{1cm}\\ \mbox{$^1$Instituto~de~Ciencias~Nucleares,~Universidad~Nacional~Aut\'onoma~de~M\'exico,}\\ \mbox{AP~70543,~Mexico~City,~Mexico}}
\affiliation{\mbox{$^2$Dipartimento~di~Fisica~and~Icra,~Universit\`a~di~Roma~“La~Sapienza”,~Roma,~Italy}}
\affiliation{\mbox{$^3$Al-Farabi~Kazakh~National~University,~Al-Farabi~av.~71,~050040~Almaty,~Kazakhstan}\\}

\date{\today}

\begin{abstract}
We explore the connection between black hole shadows, thermodynamic phase structure, and microstructure of charged and rotating black holes within General Relativity and Geometrothermodynamics. Focusing on Reissner-Nordström and Kerr solutions, we establish a criterion to select the most suitable  Geometrothermodynamic metric for a system, revealing that the first metric from enthalpy and the second from mass correctly reproduce heat capacity singularities. We show that the shadow radius encodes the same phase information as entropy and introduce Shadow–Microstructure diagrams to extract insights into stability and microscopic interaction types directly from observational bounds. Applying this framework to Sagittarius A*, we constrain the macroscopic parameters and the allowed microscopic thermodynamic phases. Our findings indicate that shadow measurements offer a novel probe of thermodynamic and microscopic aspects of black holes, enabling tests of alternative theories of gravity and  thermodynamic frameworks.\\

\end{abstract}
\maketitle
\tableofcontents
\section{Introduction}
The recent direct imaging of supermassive black holes by the Event Horizon Telescope (EHT), including M87* and Sagittarius A*, has opened an unprecedented observational window into the strong-field regime of gravity~\cite{EventHorizon1, EventHorizon2, EHT5-1,EHT5-2}. Beyond confirming General Relativity at horizon scales, these observations provide measurable shadow features that can encode not only geometric information but also the thermodynamic properties of black holes. Constraints on black hole parameters derived from shadow observational bounds have been extensively investigated in recent years across various alternative theories of gravity, making this a highly active research area~\cite{Antoniou2023, Filho_2025, LingYi, ESLAM2025, Zare_2026, Walia_2024, Vagnozzi_2023, Cao_2023, Ahmed2025, Walia_2023, Zahid2023, Sahoo, Lemos_2024, Ban2025, Zhao_2024, Puli_e_2023, G_mez_2024, Chenn2024, Tsukamoto_2024, Khodadi_2022, Shaikh_2023,2025ApJ...995..148E,2025PDU....4701734Z,2025EPJC...85.1085W,2025JHEAp..4700367A,2025A&A...693A.265E,2026PDU....5102203G,2025PDU....4701785N,2025PDU....4902017T,2025EPJC...85..878B,2025PhLB..86839812L,2026NuPhB102217212R,2025EPJC...85..973R,2025JCAP...11..069Y,2024PDU....4401501C,2024EPJC...84..136L,2024PDU....4401455W,2024PhRvD.109f4064K,bambi2019testing,vagnozzi2019hunting,allahyari2020magnetically,khodadi2024event,afrin2023tests}. In parallel, differential geometry approaches provide a powerful framework for analyzing thermodynamic systems, and related formalisms have been widely used to study phase transitions, stability, and microstructure in a wide range of black hole solutions \cite{wei2021general,Banerjee,ref20,Zangeneh_2018,Dehyadegari,Sahay,Hazarika,wei2020extended,ruppeiner2008thermodynamic,quevedo2008geometrothermodynamics,Ladino:2024ned,ladino2025phase,Larranga2011,Gogoi,larranaga2012geometric,ref54,ref39,Tharanath_2015,ZHANG2018170,Channuie:2018mkt,2025EPJC...85..785G,2018GReGr..50...20K,Sanchez:2016ger,Quevedo:2019wbz,2017GReGr..49..148H,2016ChPhB..25l0401G,2025IJGMM..2250049J,2025PDU....5002146S,Taj:2012sir,Mo:2013qhv,Quevedo_2009,Luciano_2023,Luciano_20232,2023EPJC...83..710R,2024Entrp..26..457B,Janke_2010,Quevedo_2016,Hendi_2015,Han_2012,Bravetti_2013,Quevedo:2011fk,ahmed2026shadow,Rizwan:2018ozh,Quevedo:2016swn,nojiri2021area,nojiri2022alternative,nojiri2022nonextensive,elizalde2025black}. In particular, the formalism of Geometrothermodynamics (GTD) has provided a Legendre-invariant description of thermodynamics \cite{quevedo2007geometrothermodynamics,quevedo2023unified,bravetti2017zeroth}, characterizing phase structure and microscopic interactions through curvature scalars on the equilibrium manifold. Curvature singularities indicate thermodynamic critical behavior, while the sign of the scalar curvature reflects the dominant microscopic interaction, whether attractive or repulsive. GTD has been successfully applied to systems ranging from ideal gases and van der Waals fluids~\cite{quevedo2022geometrothermodynamics,quevedo2011phase}, magnetic materials~\cite{quevedo2024geometrothermodynamic}, econophysics~\cite{2023IJGMM..2050057Q} and the Ising model~\cite{bravetti2014representation}, to black holes in various gravitational theories~\cite{quevedo2008geometrothermodynamics,Ladino:2024ned,ladino2025phase,Larranga2011,Gogoi,larranaga2012geometric,ref54,ref39,Tharanath_2015,ZHANG2018170,Channuie:2018mkt,2025EPJC...85..785G,2018GReGr..50...20K,Sanchez:2016ger,Quevedo:2019wbz,2017GReGr..49..148H,2016ChPhB..25l0401G,2025IJGMM..2250049J,2025PDU....5002146S,Taj:2012sir,Mo:2013qhv,Quevedo_2009,Luciano_2023,Luciano_20232,2023EPJC...83..710R,2024Entrp..26..457B,Janke_2010,Quevedo_2016,Hendi_2015,Han_2012,Bravetti_2013,Quevedo:2011fk,Rizwan:2018ozh,Quevedo:2016swn}, cosmology~\cite{romero2026quasi,e25060944,PhysRevD.86.063508, 2019EPJC...79..577B,2014GrCo...20..208Q}, and chemical reactions~\cite{quevedo2014geometric}.\\

Several studies in alternative theories of gravity have investigated the connection between the thermodynamic phase structure of black holes and their photon orbits and shadow features \cite{Wei_2019,ZhangGuo,Luo,GuoLi,Belhaj,WangRuppeiner,Li,Kumar,Du,Zheng,He,ref20,ref21, Du_2023, Karthik2025}. This line of research has led to the development of the shadow thermodynamics framework, which examines black hole thermodynamics through shadow properties, demonstrating that observable shadow characteristics can act as reliable probes of thermodynamic behavior. In particular, within the Ruppeiner framework, shadow observables have been shown to capture the phase structure and microstructure of anti-de Sitter (AdS) black holes \cite{WangRuppeiner,ref20,ref21}. In our previous work \cite{Ladino:2024ned,ladino2025phase}, we established an analogous correspondence for Reissner–Nordström-AdS and Kerr-AdS black holes within the GTD framework. Building on these results, we now extend the analysis to asymptotically flat charged and rotating black holes, establishing a correspondence between shadow observables and their thermodynamic and microscopic structure within GTD.\\

We focus on the Reissner–Nordström (RN) and Kerr solutions within General Relativity, analyzing their thermodynamics via the Helmholtz free energy and their GTD behavior. A central goal of this study is to clarify how the thermodynamic phase structure in GTD depends on the choice of thermodynamic potential by examining the construction of the GTD metrics and their relation to the singularities of thermodynamic response functions, particularly the heat capacities. We then reformulate the thermodynamic description in terms of black hole shadow features, investigating whether shadows can encode the same phase information typically expressed via entropy, and illustrating this through shadow thermodynamic profiles. Based on this correspondence, we introduce the concept of Shadow-Microstructure (SM) diagrams as a tool to capture thermodynamic stability and microscopic interactions directly from shadow observables. Finally, we apply this framework to Sagittarius A*, exploring how shadow measurements can constrain both macroscopic parameters and underlying microscopic behavior, while assessing their compatibility with thermodynamic stability. This paper is organized as follows. In Sec.~\ref{sec:thermo}, we review the thermodynamic phase structure of RN and Kerr black holes. In Sec.~\ref{GTDsection}, we analyze their microstructure within the GTD framework. In Sec.~\ref{sec:shadows}, we summarize the description of black hole shadows in entropy space. In Sec.~\ref{sec:sagi}, we apply the formalism to Sagittarius A* and construct the SM diagrams, identifying the observationally allowed microscopic thermodynamic phases. Finally, in Sec.~\ref{sec:conclusions}, we summarize our conclusions and discuss future perspectives.

\section{Phase Structure}\label{sec:thermo}
The phase structure of black holes depends, in general, on the number of thermodynamic degrees of freedom of the system, i. e., the number of independent variables that enter the fundamental thermodynamic equation \cite{callen1998thermodynamics}. In this work, we focus on systems with two degrees of freedom, which include static charged black holes on one side and stationary rotating black holes on the other.

\subsection{Reissner-Nordström Black Hole Thermodynamics}
The RN black hole is the charged, static solution of Einstein-Maxwell theory. Its thermodynamic behavior is characterized by a reduced set of macroscopic variables, namely the mass $M$, electric charge $Q$, entropy $S$, Hawking temperature $T$, and electric potential $U$. The four-dimensional RN black hole spacetime is given by \cite{chandrasekhar}
\begin{equation}
ds^{2}
= - f(r) dt^{2}
+ f(r)^{-1} dr^{2}
+ r^{2} d\Omega_{2}^{2},
\qquad
f(r) = 1 - \frac{2M}{r} + \frac{Q^{2}}{r^{2}},
\label{metricRN}
\end{equation}
where $d\Omega_{2}^{2}$ denotes the line element of the unit two-sphere. The mass equation follows from the condition that the lapse function $f(r)$ vanishes at the horizon. At the outer horizon $r_+$, this condition yields the mass as
\begin{equation}
M(r_+,Q) = \frac{r_+^2 + Q^2}{2 r_+},
\qquad
M(S,Q) = \frac{S + \pi Q^2}{2 \sqrt{\pi S}},
\label{fundame1}
\end{equation}

\begin{figure}[h!]
\begin{minipage}[t]{0.53\linewidth}
 \includegraphics[width=\linewidth]{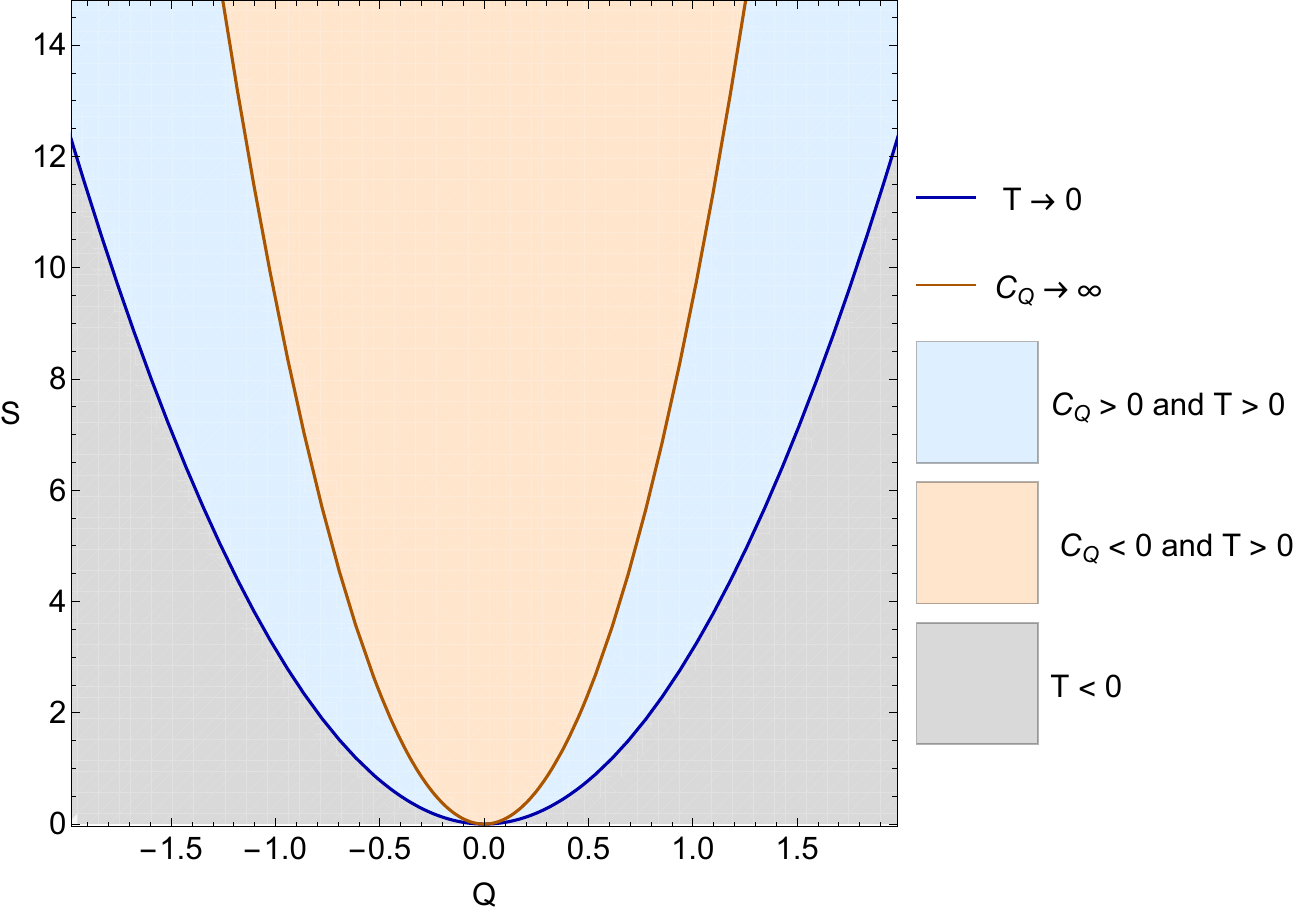}
 \\[4pt]
 (a)
\end{minipage}
\hfill
\begin{minipage}[t]{0.45\linewidth}
 \includegraphics[width=\linewidth]{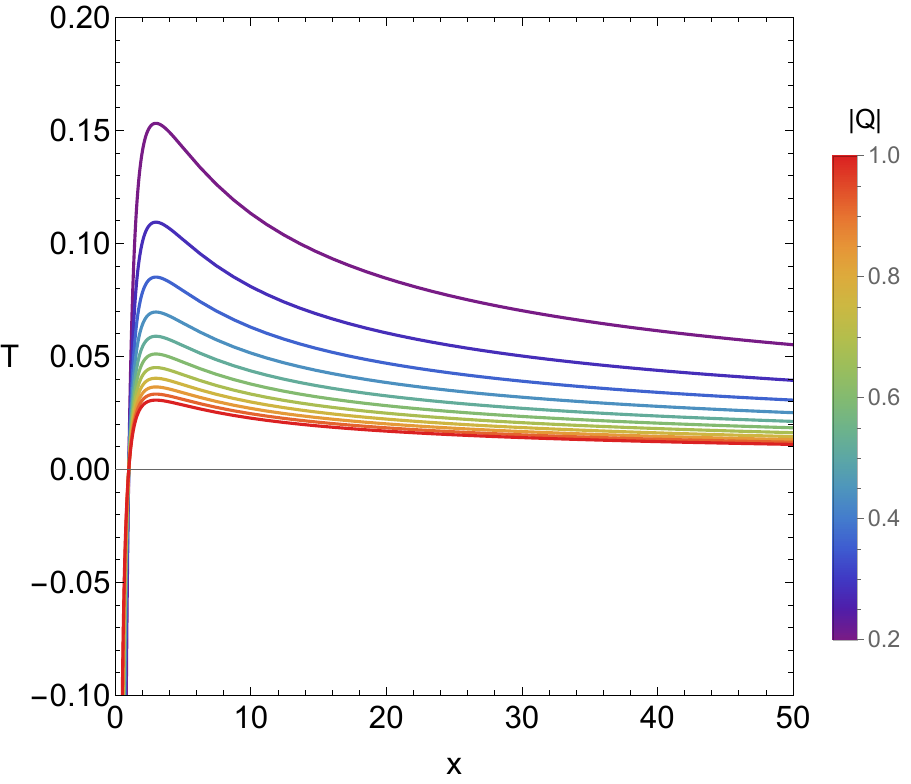}
 \\[4pt]
 (b)
\end{minipage}

\vspace{0.5cm}

\begin{minipage}[t]{0.47\linewidth}
 \includegraphics[width=\linewidth]{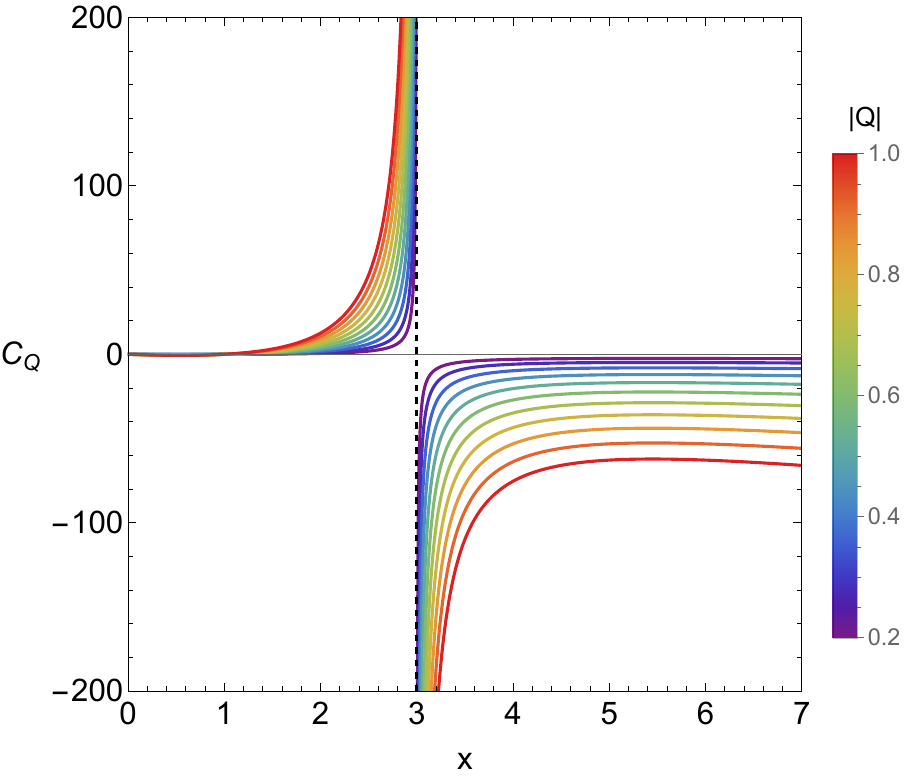}
 \\[4pt]
 (c)
\end{minipage}
\hfill
\begin{minipage}[t]{0.46\linewidth}
 \includegraphics[width=\linewidth]{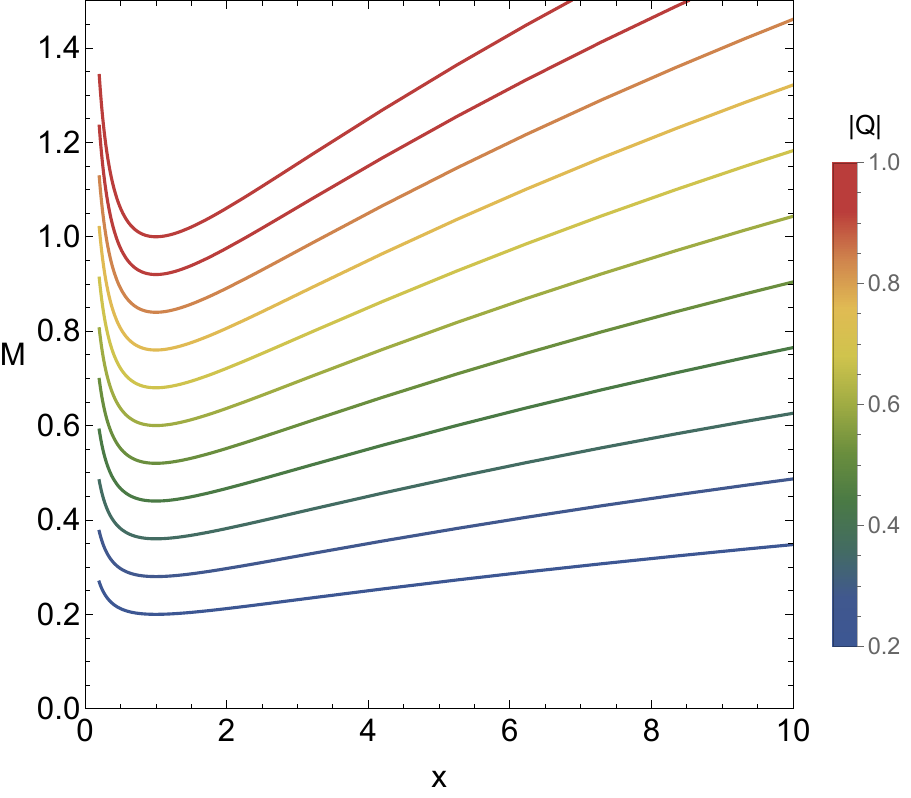}
 \\[4pt]
 (d)
\end{minipage}
\caption{
(a) Black hole existence lies above the blue parabola $S = \pi Q^2$; 
the orange parabola $S_m = 3\pi Q^2$ is the critical curve of $C_Q$. 
Local thermodynamic stability at fixed charge occurs between the two curves. 
(b) Temperature, (c) heat capacity at constant $Q$, and (d) mass, 
all expressed in terms of $x \equiv S/\pi Q^2$.
}
\label{Temperature RN}
\end{figure}
\begin{figure}[H]
\begin{minipage}[t]{0.48\linewidth}
 \centering
 \hspace{-4cm} \includegraphics[width=1\linewidth]{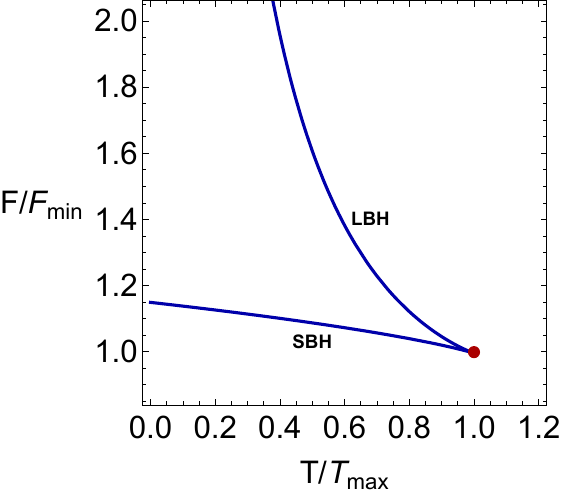}
 \hspace{-5cm}(a)
\end{minipage}%
\hfill%
\begin{minipage}[t]{0.44\linewidth}
 \centering
 \hspace{-4cm} \includegraphics[width=1\linewidth]
 {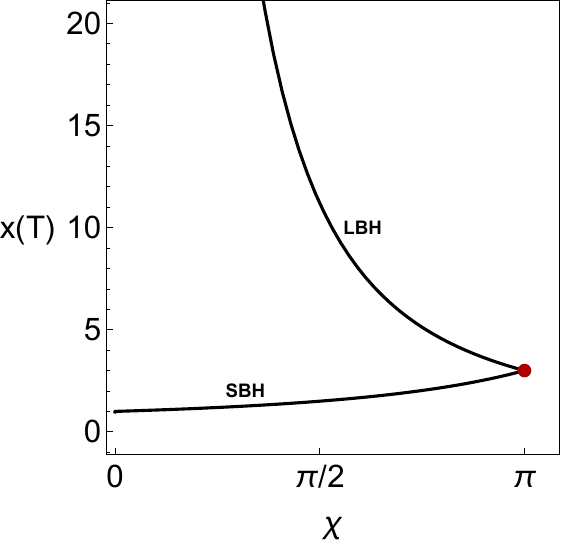}
\hspace{-5cm}(b)
\end{minipage}%
\hfill%
\caption{(a) Helmholtz free energy of the RN black hole; (b) parametrized entropy branches in terms of the reduced variable $x \equiv S/\pi Q^2$. The limits $\chi \to 0$ and $\chi \to \pi$ correspond to $x \to 1$ ($T=0$) and $x \to 3$ ($T = T_{\max}$), respectively.}

\label{Free energy RN}
\end{figure}

\noindent
where in the second expression, we have rewritten the mass in terms of the entropy $S=\pi r_+^2$, which defines the fundamental thermodynamic equation. Moreover, in the canonical ensemble, which allows the electric charge $Q$ to fluctuate \cite{caldarelli2000thermodynamics}, the thermodynamic first law takes the form
\begin{equation}
\label{firstlaw}
dM = TdS + UdQ ,
\end{equation}
where $U$ is interpreted as the electric potential.
Accordingly, the thermodynamic quantities can be expressed as equations of state derived from Eq.~\eqref{fundame1}, namely,
\begin{equation}
\label{tempe}
T(S,Q)=\left(\frac{\partial M}{\partial S}\right)_{Q}
=\frac{S-\pi Q^{2}}{4\sqrt{\pi}S^{3/2}},
\qquad
U(S,Q)=\left(\frac{\partial M}{\partial Q}\right)_{S}
=\frac{\sqrt{\pi}Q}{\sqrt{S}} .
\end{equation}
The temperature reaches a maximum value at
\begin{equation}
S_{m} = 3\pi Q^2,\qquad T_{\text{max}}=\frac{1}{6\sqrt{3}\pi |Q|},\label{maxtemRN}
\end{equation}
where the system satisfies the universal relation $Q/M=\sqrt{3}/2$. Additionally, the heat capacity at constant $Q$ is given by
\begin{align}
C_{Q}&=T\left(\frac{\partial S}{\partial T}\right)_{Q}=\frac{2 S (S - \pi Q^2)}{3 \pi Q^2 - S}.
\end{align}
The black hole solution exists for entropies \( S > \pi Q^2 \) and presents a Davies point \cite{davies1978thermodynamics} at \( S = S_m \). In the interval \( \pi Q^2 < S < 3\pi Q^2 \), it remains locally thermodynamically stable, with positive heat capacity (see Fig.~\ref{Temperature RN}(a)). Introducing the dimensionless parameter \( x \equiv S/(\pi Q^2) \) conveniently sets the scale governing the thermodynamic behavior, as illustrated in Fig.~\ref{Temperature RN}(b)–(d). In terms of \(x\), the mass, temperature, and heat capacity can be written as
\begin{equation}
\label{reducedtempe}
M
=
\frac{|Q|}{2}
\left(\frac{x+1}{\sqrt{x}}
\right), \quad T =\frac{1}{4\pi |Q|}
\left(\frac{x-1}{x^{3/2}}\right),
\quad 
C_Q =2\pi x Q^2\left(\frac{ x - 1}{3 - x}\right).
\end{equation}
To investigate the global thermodynamic behavior of the black hole, we use the Helmholtz free energy, defined as
\begin{align}
F \equiv M-TS= \frac{|Q|}{4}\left(\frac{x+3}{\sqrt{x}}\right).
\end{align}
Physically, introducing \(x\) makes explicit that the electric charge \(Q\) appears only as an overall multiplicative factor in the thermodynamic quantities. Hence, varying \(Q\) merely rescales the system without altering its thermodynamic structure or phase behavior. Moreover, from the temperature relation one can solve for \(x(T)\) by defining the auxiliary variables \(u \equiv \sqrt{x}\) and \(y \equiv 4\pi |Q| T\), yielding
\begin{equation}
    y u^3 - u^2 + 1 = 0.\label{RN branches}
\end{equation}
A convenient trigonometric parametrization is obtained by introducing an angle \(\chi\) such that
\begin{equation}
    y(\chi)=
\sqrt{\frac{2}{27}\bigl(1-\cos\chi\bigr)}  , \qquad 0 \leq \chi \leq \pi . \label{y relation}
\end{equation}
The corresponding solutions can then be written as
\begin{equation}
    x_{i}(\chi)=\frac{1}{9y^2}\left(1+2\cos\frac{\chi - 2\pi i}{3}\right)^2,
    \qquad 
    i=0,1,2. \label{x_branches}
\end{equation}
For temperatures in the range \(0\le T\le T_{{max}}\), we have \(0\le\chi\le\pi\), and the roots satisfy
\(x_0>x_1>0>x_2\). Thus, in direct analogy with the AdS black hole case, the roots $x_0$ and $x_1$ correspond to the large (LBH) and small black (SBH) hole branches, respectively. The root $x_2$ is unphysical, as it yields a negative temperature. As a result, the Helmholtz free energy exhibits two distinct thermodynamic branches, as shown in Fig. \ref{Free energy RN}
The Helmholtz free energy satisfies the thermodynamic identity
\begin{equation}
\left(\frac{\partial F}{\partial T}\right)_{Q} = -S ,
\end{equation}
which remains finite for the RN black hole. However, when the free energy is expressed as a function of the temperature along the equation of state, $F=F(x(T))$, the relevant quantity is the total derivative,
\begin{equation}
\frac{dF}{dT}
=
\left(\frac{\partial F}{\partial x}\right)_{T}
\frac{dx}{dT}.
\end{equation}
At the maximum temperature, \( dT/dx = 0 \), so that \( dx/dT \) diverges, leading to a divergence of \( dF/dT \). This Davies point behavior signals a spinodal instability rather than a genuine phase transition. The same conclusion was presented in \cite{wu2025extended} using the thermodynamic topology formalism, where RN lacks critical points. A qualitatively different situation arises when the black hole configuration involves an additional length scale, as in AdS spacetimes, where the curvature radius acts as an effective thermodynamic parameter. This extra scale enables the coexistence of distinct phases and, consequently, the emergence of phase transitions \cite{Ladino:2024ned,ladino2025phase}.

\subsection{Kerr Black Hole Thermodynamics}
\begin{figure}[H]
\begin{minipage}[t]{0.5\linewidth}
 \centering
 \hspace{-4cm} \includegraphics[width=1\linewidth]{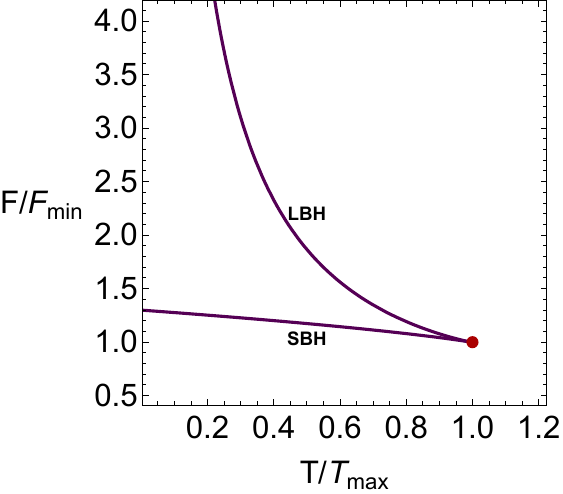}
 \hspace{-5cm}(a)
\end{minipage}%
\hfill%
\begin{minipage}[t]{0.464\linewidth}
 \centering
 \hspace{-4cm} \includegraphics[width=1\linewidth]
 {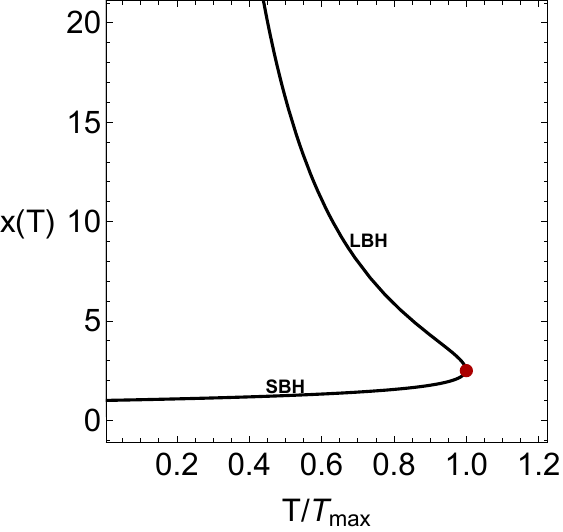}
\hspace{-5cm}(b)
\end{minipage}%
\hfill%
\caption{(a) Helmholtz free energy of the Kerr black hole; (b) entropy branches in terms of  $x \equiv S/2\pi|J|$.}
\label{Free energy Kerr}
\end{figure}

The Kerr solution is the stationary, axisymmetric vacuum solution of Einstein's field equations that describes the spacetime geometry outside a rotating, uncharged black hole. It is fully characterized by the mass $M$ and the angular momentum $J$ of the black hole. The Kerr metric in Boyer-Lindquist coordinates ($t,r,\theta,\phi$) is given by \cite{Hioki_2009, chandrasekhar, Li_2014}
\begin{equation}
    ds^2 = - \left( 1 - \frac{2 M r}{\Sigma} \right) dt^2
- \frac{4 M a r \sin^2\theta}{\Sigma}  dt  d\phi
+ \frac{\Sigma}{\Delta}  dr^2
+ \Sigma  d\theta^2  + \left( r^2 + a^2 + \frac{2 M a^2 r \sin^2\theta}{\Sigma} \right) \sin^2\theta  d\phi^2,\label{metric_Kerr}
\end{equation}
where
\begin{equation}
 \Sigma = r^2 + a^2 \cos^2\theta, \qquad
\Delta = r^2 - 2 M r + a^2.  
\end{equation}
The mass of the Kerr black hole in terms of the outer horizon radius $r_+$ and the rotation parameter $a = J/M$ is obtained from the condition $\Delta(r_+) = 0$, yielding
\begin{equation}
M = \frac{r_+^2 + a^2}{2 r_+},
\qquad
M(S,J) = \sqrt{\frac{4\pi^2 J^2 + S^2}{4 \pi S}},
\label{fundamenta Kerr}
\end{equation}
\begin{figure}[h]
\centering
\begin{minipage}[t]{0.54\linewidth}
 \centering
 \includegraphics[width=\linewidth]{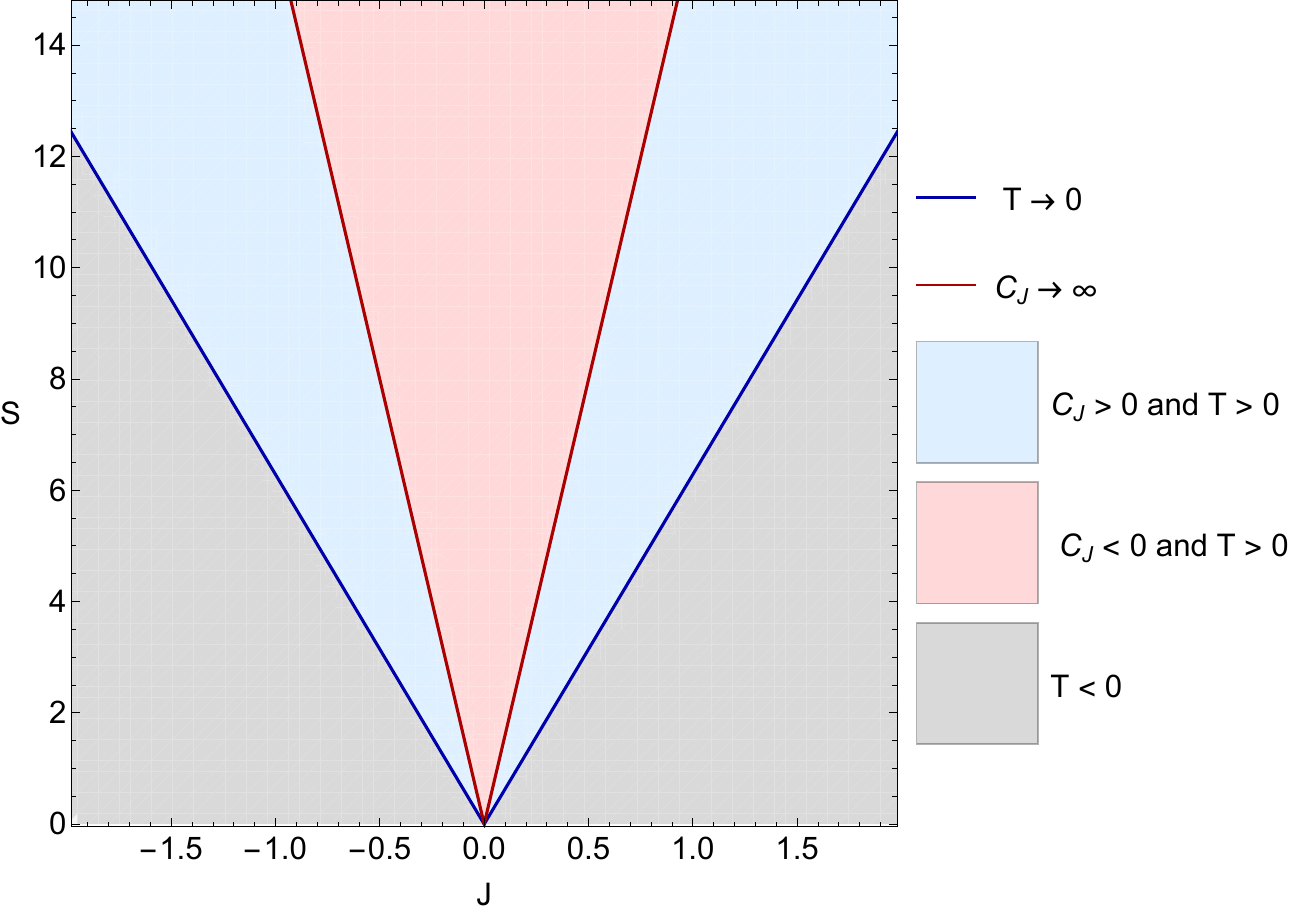}
 \\[4pt]
 (a)
\end{minipage}
\hfill
\begin{minipage}[t]{0.45\linewidth}
 \centering
 \includegraphics[width=\linewidth]{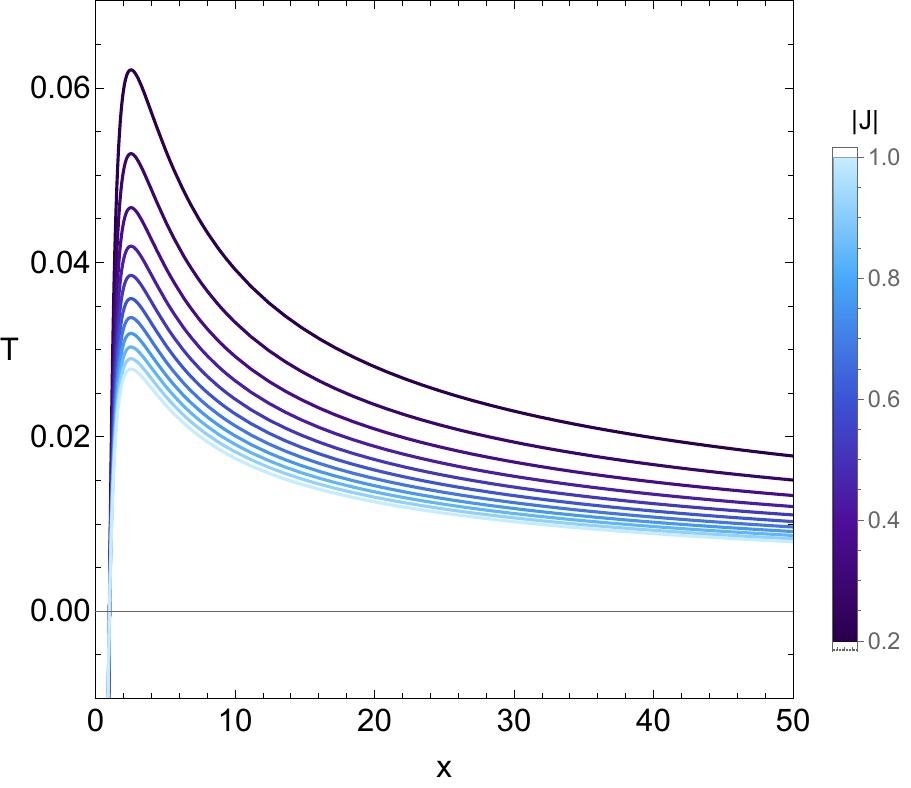}
 \\[4pt]
 (b)
\end{minipage}

\vspace{0.5cm}

\begin{minipage}[t]{0.47\linewidth}
 \centering
 \includegraphics[width=\linewidth]{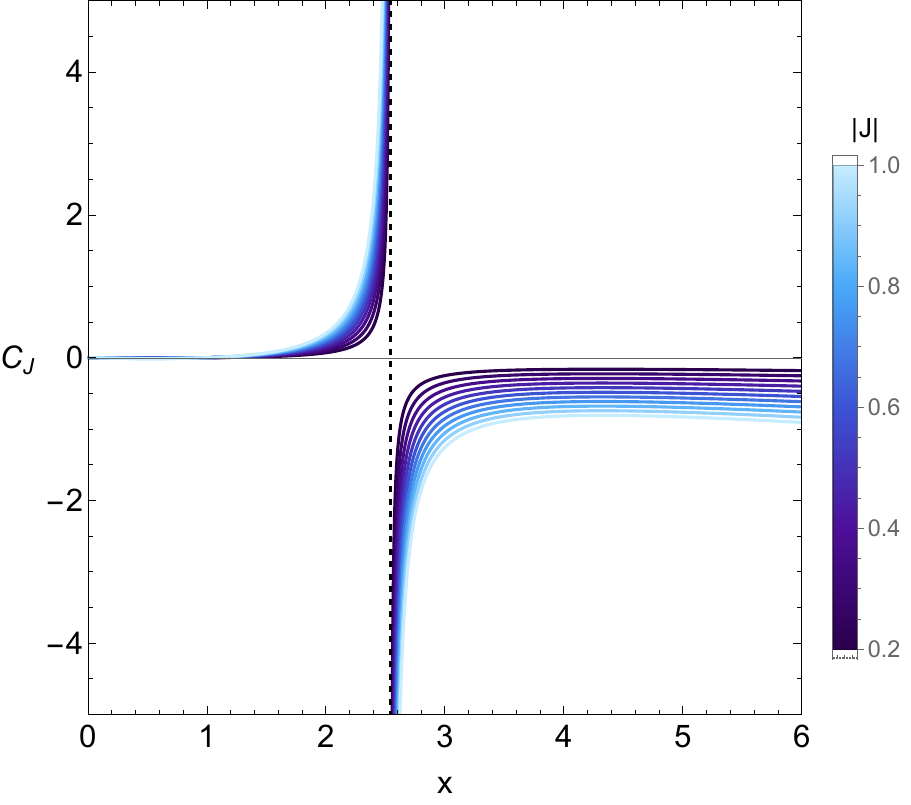}
 \\[4pt]
 (c)
\end{minipage}
\hfill
\begin{minipage}[t]{0.47\linewidth}
 \centering
 \includegraphics[width=\linewidth]{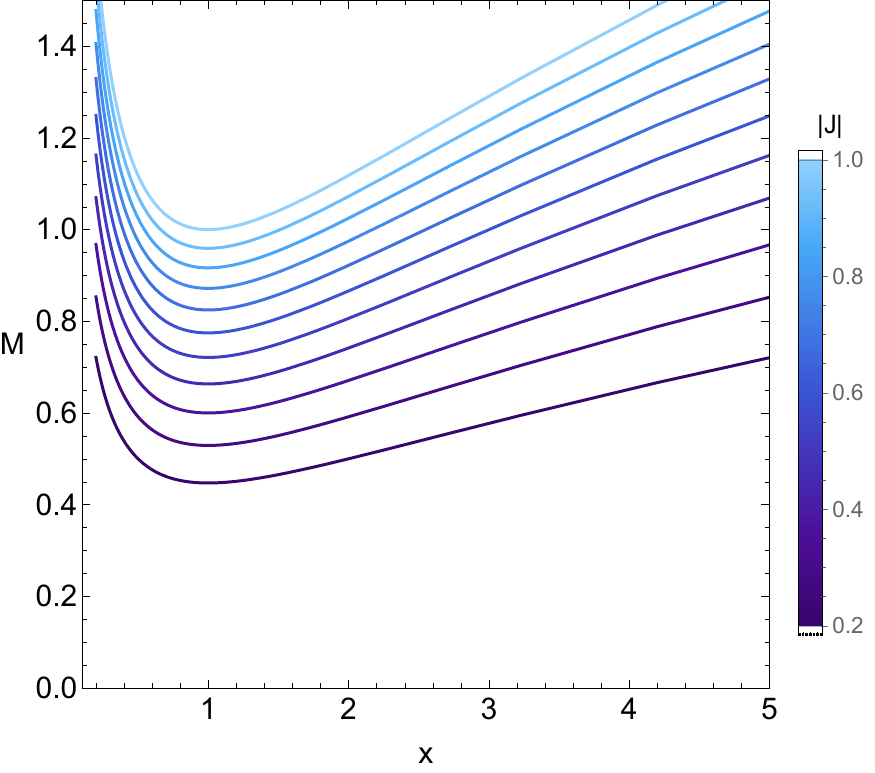}
 \\[4pt]
 (d)
\end{minipage}
\caption{
(a) Black hole existence lies above the blue curve $S = 2\pi |J|$; the red curve $S = S_{m}$ is the critical curve of $C_J$. Local thermodynamic stability at fixed $J$ occurs between the two curves. (a) Temperature, (b) heat capacity at constant $J$, and (c) mass, all expressed in terms of $x \equiv S/2\pi|J|$.}
\label{Temperature Kerr}
\end{figure}
where the Kerr entropy $S = \pi \left(r_+^2 + a^2\right)$ has been used to express the mass in terms of $(S,J)$, and the existence of the event horizon, $r_+ = M + \sqrt{M^2 - a^2}$, imposes the condition $M \geq \sqrt{|J|}$.
In the energy representation, the first law of black hole thermodynamics reads
\begin{equation}
dM = T dS + \Omega dJ,
\end{equation}
where $\Omega$ denotes the angular velocity of the event horizon. In this representation, the thermodynamic behavior of the solution can be characterized through two equations of state
\begin{equation}
T(S,J) = \left(\frac{\partial M}{\partial S}\right)_{J} 
= \frac{S^2 - 4 \pi^2 J^2}{4 \sqrt{\pi S^3 (4 \pi^2 J^2 + S^2)}}, 
\qquad 
\Omega(S,J) =\left(\frac{\partial M}{\partial S}\right)_{S}= \frac{2 \pi^{3/2} J}{\sqrt{S \left(4 \pi^2 J^2+S^2\right)}}.
\end{equation}
At fixed $J$, the temperature as a function of the entropy exhibits a maximum at
\begin{equation}
S_{m}=
2\pi|J|(3+2\sqrt{3})^{1/2}, \qquad T_{\text{max}}=\frac{1}{2\sqrt{2}\pi(135+78\sqrt{3})^{1/4}|J|^{1/2}}\approx0.028|J|^{-1/2},\label{maxtemKerr}
\end{equation}
at which the system satisfies the universal relation $\sqrt{|J|}/M
= \left(1+2/\sqrt{3}\right)^{-1/4}
\approx 0.825$. The heat capacity at constant $J$ is given by
\begin{align}
C_{J}&=T\left(\frac{\partial S}{\partial T}\right)_{J}=\frac{2S\left(2\pi J - S\right)\left(2\pi J + S\right)
\left(4\pi^2 J^2 + S^2\right)}
{S^4-48\pi^4 J^4- 24\pi^2 J^2 S^2 }.
\label{CJ_kerr}
\end{align}
The Kerr black hole exists for entropies \( S > 2\pi |J| \) and features a Davies point \cite{davies1978thermodynamics} at \( S = S_m \). It is locally thermodynamically stable under fixed angular momentum, as indicated by a positive heat capacity (see Fig.~\ref{Temperature Kerr}(a)). As in the RN case, introducing the dimensionless parameter \( x \equiv S/(2\pi |J|) \) is convenient, in terms of which the mass, temperature, and heat capacity can be expressed as
\begin{equation}
\label{reducedtempe_kerr}
M
=|J|^{1/2}
\left(\frac{x^2+1}{2x}
\right)^{1/2}, \quad T =\frac{1}{4\sqrt{2}\pi |J|^{1/2}}
\left(\frac{x^2-1}{x^{3/2}\sqrt{x^2+1}}\right),
\quad 
C_J =4\pi x |J| \left(\frac{ 1-x^4 }{x^4-6x^2-3}\right).
\end{equation}
Figure~\ref{Temperature Kerr} shows that the thermodynamic parameters display the same qualitative behavior as in the RN case. As $x \to \sqrt{3+2\sqrt{3}}$, the temperature attains its maximum value $T_{\text{max}}$, and the heat capacity diverges, indicating the presence of a Davies point. Furthermore, the free energy for the Kerr black hole is
\begin{align}
F\equiv M-TS= |J|^{1/2}\left(\frac{x^2+3}{2\sqrt{2x(x^2+1)}}\right).
\end{align}
The temperature relation can be written as
\begin{equation}
    2y^{2}x^{3}(x^{2}+1)=(x^{2}-1)^{2},\label{kerr branches}
\end{equation}
where $y\equiv 4\pi\sqrt{|J|}T$. The above relation defines $x$ implicitly as a function of the temperature $T$. In contrast to the RN solution Eq.~\eqref{RN branches}, for which  can be solved analytically, Eq.~\eqref{kerr branches} does not admit a closed-form solution. Nevertheless, a numerical analysis can be performed, as illustrated in Fig.~\ref{Free energy Kerr}.
Figure~\ref{Free energy Kerr} shows the phase structure of the Kerr black hole. As in the RN case, for $0<T<T_{\max}$ two physical branches are present, corresponding to a stable SBH and an unstable LBH, which merge at $T_{\max}$. In addition, a third branch with $T<0$ appears, analogous to that found in the
RN solution; however, this branch is  unphysical and is therefore excluded
from the analysis. Once again, this phase-structure behavior at the Davies point is more closely
associated with a spinodal instability rather than with a genuine phase transition.
\section{Geometrothermodynamics}\label{GTDsection}
GTD, introduced in \cite{quevedo2007geometrothermodynamics}, is a geometric framework for studying the thermodynamic properties of physical systems. In GTD, singularities of the Ricci scalar coincide with thermodynamic critical points or phase transitions, such as the liquid–gas critical point in fluids or the small–large black hole transition in AdS space, while its sign indicates whether repulsive interactions dominate when positive, attractive interactions dominate when negative, or the system is non-interacting when zero, as in an ideal gas ~\cite{quevedo2023unified,Ladino:2024ned,ladino2025phase}. A key feature of GTD, absent in other geometric approaches such as Ruppeiner thermodynamics~\cite{ruppeiner1979thermodynamics,ruppeiner1981application}, is its invariance under Legendre transformations, which guarantees that the thermodynamic description is independent of the chosen potential~\cite{callen1998thermodynamics}. This invariance is implemented by introducing an auxiliary $(2n+1)$-dimensional manifold $\mathcal{T}$, coordinatized by $Z^A=\{\Phi,E^a,I_a\}$, where $n$ is the number of thermodynamic degrees of freedom and $\Phi$ the thermodynamic potential. While $E^a$ and $I_a$ naturally correspond to extensive and intensive variables in homogeneous systems, this identification becomes subtler for quasi-homogeneous systems such as black holes~\cite{quevedo2023unified,bravetti2017zeroth}. Currently, three Legendre-invariant metrics are known, given by
\begin{align}
    g^I_{ab}&=\nu_\Phi \Phi \delta^c_a\frac{\partial^2 \Phi}{\partial E^b \partial E^c}, \label{g1}\\
    g^{II}_{ab}&=\nu_\Phi \Phi \eta^c_a\frac{\partial^2 \Phi}{\partial E^b \partial E^c},  \label{g2}\\
    g^{III}&=\sum_{a=1}^{n}\nu_a\left(\delta_{ad}E^d\frac{\partial \Phi}{\partial E^a}\right)\delta^{ab}\frac{\partial^2 \Phi}{\partial E^b \partial E^c} dE^a dE^c,  \label{g3}
\end{align}
 where \( \delta_{cd} = \text{diag}(1, 1, \ldots, 1)\), \( \eta_{cd} = \text{diag}(-1, 1, \ldots, 1)\), \( \nu_a \) are the quasi-homogeneity coefficients that define the scaling properties of the fundamental equation, and \( \nu_\Phi \) is the quasi-homogeneity degree of the potential \( \Phi \)~\cite{quevedo2023unified}. In GTD, thermodynamic states are represented as points in an \( n \)-dimensional subspace of \( \mathcal{T} \), known as the equilibrium space \( \mathcal{E} \). The first law of thermodynamics is naturally satisfied on \( \mathcal{E} \), and the coordinates \( Z^A \) become functions of the variables \( E^a \), that is $ Z^A(E^a) = \{ \Phi(E^a), E^a, I_a(E^a) \}$, where \( \Phi = \Phi(E^a) \)  represents the fundamental equation of the thermodynamic system~\cite{callen1998thermodynamics} and $I_a=\partial \Phi/\partial E^a $ the dual variables. In the present work, we consider a two-dimensional equilibrium manifold. The general structure of the corresponding independent curvature scalars of the GTD metrics has been analyzed in \cite{quevedo2023unified,Ladino:2024ned}, where it was concluded that, if we demand the singularities of $g^{III}$ to be compatible with those of $g^{I}$ and $g^{II}$, then all singularities are determined by the zeros of the second-order derivatives of $\Phi$, namely 
\begin{align}
    I&:\ \Phi_{,11} \Phi_{,22} - \big(\Phi_{,12}\big)^2 = 0, \label{c1}\\
   II&:\ \Phi_{,11} \Phi_{,22} = 0, \\
  III&:\ \Phi_{,12} = 0, \label{c3}
\end{align}
which are associated with the onset of thermodynamic instabilities. This becomes clearer when the singularity conditions are rewritten in terms of the system’s thermodynamic response functions (see Appendix B in Ref.~\cite{Ladino:2024ned}). However, the resulting phase transition scheme depends on the thermodynamic potential used to evaluate conditions Eqs.~\eqref{c1}–\eqref{c3}. For black holes, stability properties are known to depend on the statistical ensemble~\cite{caldarelli2000thermodynamics}, although in the thermodynamic limit different ensembles are related by Legendre transformations~\cite{quevedo2014ensemble}. Consequently, the phase structure may vary with the chosen potential. Here, we establish a criterion to identify the GTD metric most appropriate for a given system. Assuming the validity of the quasi-homogeneous Euler identity \cite{romero2024extended1}, our results for a two-dimensional equilibrium manifold are summarized in Table~\ref{Table I}. 
\begin{table}[h]
\centering
\setlength{\tabcolsep}{4pt}
\renewcommand{\arraystretch}{1.05}
\begin{tabular}{c|c}
\hline
\textbf{Potential} & \textbf{Curvature Correspondence} \\
\hline
$M(S,E_1)$ &
$\begin{array}{l}
\mathcal{R}^{I}_{N} \sim (C_{I_1}\kappa_S)/T \\
\mathcal{R}^{II}_{N} \sim (C_{E_1}\kappa_S)/T \\
\mathcal{R}^{III} \sim \alpha_S
\end{array}$ \\
\hline
$H(S,I_1)$ &
$\begin{array}{l}
\mathcal{R}^{I}_{N} \sim C_{E_1}/(T\kappa_S) \\
\mathcal{R}^{II}_{N} \sim C_{I_1}/(T\kappa_S) \\
\mathcal{R}^{III} \sim \hat{\alpha}_S
\end{array}$ \\
\hline
$F(T,E_1)$ &
$\begin{array}{l}
\mathcal{R}^{I}_{N} \sim C_{I_1}/(T\kappa_T) \\
\mathcal{R}^{II}_{N} \sim (T\kappa_T)/C_{E_1} \\
\mathcal{R}^{III} \sim (T\alpha_S)/C_{E_1}
\end{array}$ \\
\hline
$G(T,I_1)$ &
$\begin{array}{l}
\mathcal{R}^{I}_{N} \sim (C_{E_1}\kappa_T)/T \\
\mathcal{R}^{II}_{N} \sim T/(C_{I_1}\kappa_T) \\
\mathcal{R}^{III} \sim (T\hat{\alpha}_S)/C_{E_1}
\end{array}$ \\
\hline
$S(M,E_1)$ &
$\begin{array}{l}
\mathcal{R}^{I}_{N} \sim T C_{I_1}\kappa_S \\
\mathcal{R}^{II}_{N} \sim (T C_{E_1}\kappa_M \alpha_M)/(I_1\kappa_M - T\alpha_M) \\
\mathcal{R}^{III} \sim T\alpha_M
\end{array}$ \\
\hline
\end{tabular}
\caption{
Correspondence between normalized GTD curvature scalars and singularities/zeros of response functions for different thermodynamic potentials, with $\mathcal{R}_N=\Phi^3\mathcal{R}(q_1,q_2)$.
}
\label{Table I}
\end{table}

The response functions have been defined using the Poisson bracket notation in a thermodynamic coordinate space $(q_1, q_2)$, as
\begin{align}
C_{x}&=T\left(\frac{\partial S}{\partial T}\right)_{x}=T\frac{\{S,x\}_{q_1,q_2}}{\{T,x\}_{q_1,q_2}}, \quad
\kappa_{x}=\left(\frac{\partial E_1}{\partial I_1}\right)_{x}=\frac{\{E_1,x\}_{q_1,q_2}}{\{I_1,x\}_{q_1,q_2}},\\
\alpha_{x}&=\left(\frac{\partial E_1}{\partial T}\right)_{x}=\frac{\{E_1,x\}_{q_1,q_2}}{\{T,x\}_{q_1,q_2}}, \quad
\hat{\alpha}_{x}=\left(\frac{\partial I_1}{\partial T}\right)_{x}=\frac{\{I_1,x\}_{q_1,q_2}}{\{T,x\}_{q_1,q_2}}.
\end{align}
$C_{x}$, $\kappa_{x}$, and $\alpha_{x}$ denote the generalized heat capacity, compressibility, and coefficient of thermal expansion, respectively, evaluated at a fixed thermodynamic parameter $x$ \cite{romero2024extended1}. Although Table~\ref{Table I} can be generalized to higher-dimensional equilibrium manifolds, RN and Kerr black holes are described by two-dimensional ones. From Table~\ref{Table I} we observe that, regardless of the chosen thermodynamic potential, the divergences in $\mathcal{R^{II}}$ always correspond to singularities (or zeros) of the response functions in the associated ensemble, while $\mathcal{R^{I}}$ consistently aligns with singularities (or zeros) of the heat capacity in the dual representation. Interestingly, the divergences of the Ruppeiner metric follow the same pattern as those of $g^{I}$, since they correspond to singularities of the heat capacity in the dual representation~\cite{mansoori2015hessian}. It is important to stress that only $\mathcal{R}^{I}$ is able to predict the singularities associated with extremal states ($T=0$). These divergences may be interpreted as signaling a transition between extremal configurations and naked singularities. In contrast, $\mathcal{R}^{II}$ does not capture these divergences. The reason lies in the properties of the Poisson bracket in the $(q_1,q_2)$ coordinate space, which implies that the ratio $C_{q_2} \kappa_{q_1}/T$ is independent of $T$. Thus, the extremal limit $T \to 0$ does not generate additional singularities in $\mathcal{R}^{II}$. Moreover, it is worth highlighting that the conditions that accurately reproduce the singularities of the heat capacity $C_{E_1}$ arise in the metrics $g^{I}$ and $g^{II}$ as
\begin{align}
I: \quad 
& H_{,SS} H_{,I_1 I_1} - \left( H_{,S I_1} \right)^2 = 0, \label{sing con1}\\[4pt]
II: \quad 
& M_{,SS} M_{,E_1 E_1} = 0. \label{sing con2}
\end{align}
This result provides the main justification for focusing on $g^{I}$, constructed from $H(S,I_1)$, and on $g^{II}$, constructed from $M(S,E_1)$, in the forthcoming analysis, as they produce the correct phase structure indicated by the singularities of the heat capacity $C_{E_1}$. We now apply these to the RN and Kerr black holes.

\subsection{Reissner-Nordström Microstructure}
\begin{figure}[h]
\begin{minipage}[t]{0.34\linewidth}
    \centering
    \includegraphics[width=\linewidth,height=8cm,keepaspectratio]{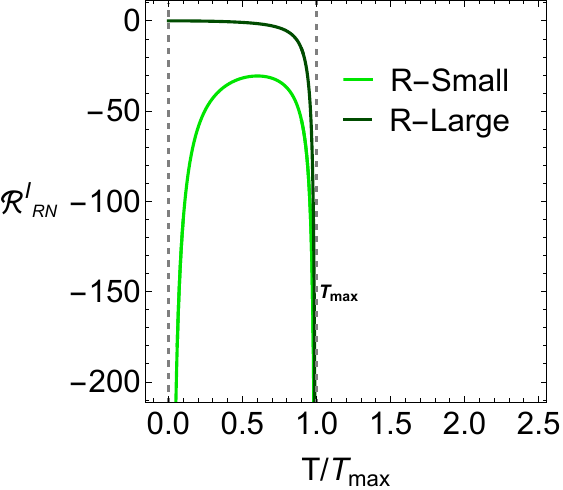}\\
    (a)
\end{minipage}
\hfill
\begin{minipage}[t]{0.32\linewidth}
    \centering
    \includegraphics[width=\linewidth,height=8cm,keepaspectratio]{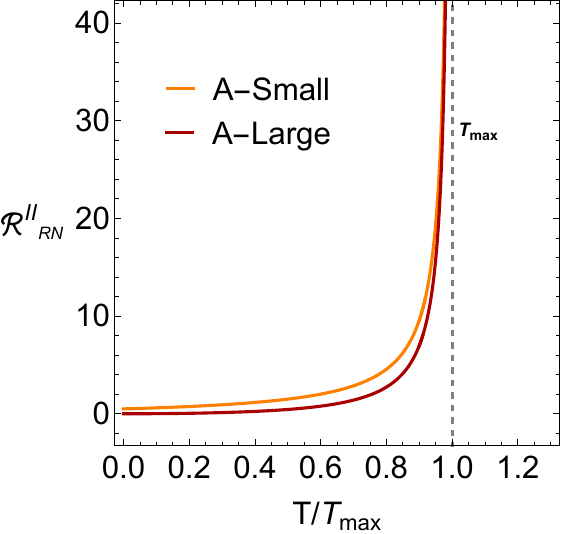}\\
    (b)
\end{minipage}
\hfill
\begin{minipage}[t]{0.31\linewidth}
    \centering
    \includegraphics[width=\linewidth,height=8cm,keepaspectratio]{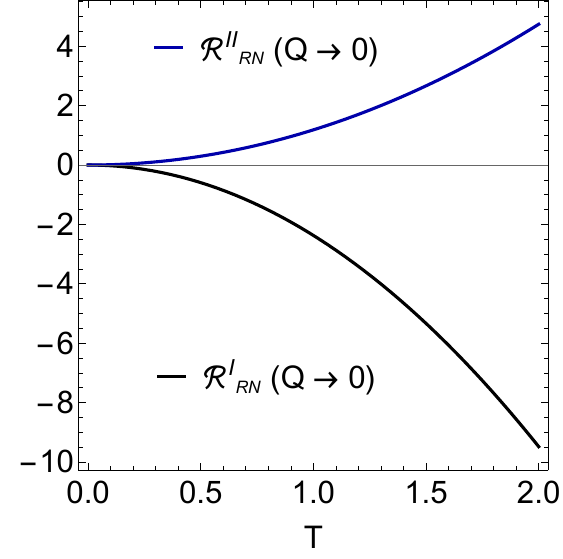}\\
    (c)
\end{minipage}

\caption{GTD scalars of the RN black hole as functions of the temperature.
(a) $\mathcal{R}^{I}$, (b) $\mathcal{R}^{II}$, and (c) comparison of both scalars in the Schwarzschild limit $Q \to 0$. Here, $R$ and $A$ stand for repulsive and attractive interactions, respectively.}
\label{RN_scalarR2_microstructure}
\end{figure}

First, we show for the RN solution the correspondence between the unified conditions for curvature singularities, Eqs.~\eqref{c1}--\eqref{c3}, and the divergence of the response functions listed in Table~\ref{Table I} using the energy representation. 
By performing a rescaling $\lambda$ of the $(S,Q)$ variables, it is straightforward to see that Eq.~\eqref{fundame1} is a quasi-homogeneous function of arbitrary degree $\nu_M$, i.e.,
\begin{equation}
M(\lambda^{\nu_S} S,\lambda^{\nu_Q} Q)=\lambda^{\nu_M} M(S,Q),
\end{equation}
provided that the condition
\begin{equation}
    \nu_Q=\frac{1}{2}\nu_S,\quad \nu_S= 2\nu_M, \label{escal}
\end{equation}
is satisfied. With this condition, the Euler identity $\nu_S T S + \nu_Q  UQ = \nu_M M$ is satisfied, and consequently, we can write the quasi-homogeneous GTD line elements, Eqs.~(\ref{g1})--(\ref{g3}), as
\begin{align}
g^I &= \nu_M M 
\left(
\frac{3 \pi Q^2 - S}{8 \sqrt{\pi} S^{5/2}}  dS^2
- \frac{Q \sqrt{\pi}}{S^{3/2}}  dS  dQ
+ \frac{\sqrt{\pi}}{\sqrt{S}}  dQ^2
\right), \\[2mm]
g^{II} &= \nu_M M 
\left(
- \frac{3 \pi Q^2 - S}{8 \sqrt{\pi} S^{5/2}}  dS^2
+ \frac{\sqrt{\pi}}{\sqrt{S}}  dQ^2
\right), \label{g2_energy}\\[1mm]
g^{III} &= \nu_M \left(
\frac{(3 \pi Q^2 - S)(- \pi Q^2 + S)}{16 \pi S^3}  dS^2
- \frac{ Q (\pi Q^2 + S)}{4 S^2}  dS  dQ
+ \frac{\pi Q^2}{S}  dQ^2
\right).
\end{align}
where we have used the Euler identity, and the relationships between the quasi-homogeneity coefficients 
Eq. (\ref{escal}). Accordingly, in light of Eqs.~\eqref{c1}--\eqref{c3} and Table~\ref{Table I}, the singularities are determined by the conditions
  \begin{align}
      I&:M_{,SS} M_{,QQ}- (M _{,SQ})^2=\frac{T}{C_U \times \kappa_S}=\frac{\sqrt{\pi}}{2S^{3/2}}T=0, \label{c1schw}\\
       II&:M_{,SS} M _{,QQ}=\frac{T}{C_Q \times \kappa_S}=
\frac{S-3\pi Q^2}{8 S^3} = 
       0,\\
       III&: M _{,SQ}= \frac{1}{\alpha_S}=\frac{\sqrt{\pi}Q}{2S^{3/2}} =  0.\label{c3schw}\
\end{align}
For physical configurations with $S\neq 0$ and $ Q \neq 0$, conditions~$I$ and~$III$ are generally not fulfilled. In contrast, singularity~$II$ at the Davies point $S_m = 3\pi Q^2$ shows that, for any fixed charge, there exists a positive entropy where the curvature diverges. Moreover, as noted above, to fully reproduce the singularities of $g^{I}$ through the heat capacity $C_Q$, the latter must be computed from the enthalpy potential
\begin{equation}
    H(S,U)\equiv M-UQ=- \frac{\sqrt{S}  (-1 + U^2)}{2 \sqrt{\pi}}.\label{RN-enthalpy}
\end{equation}
In this representation, the first law of black hole thermodynamics can be expressed as
\begin{equation}
    dH = T dS - Q dU,
\end{equation}  
where the temperature \(T\) and electric charge \(Q\) play the role of equations of state, given explicitly by
\begin{equation}
\label{tempe_grand canonical_RN}
T(S,U) = \left(\frac{\partial H}{\partial S}\right)_{U} 
= \frac{1 - U^2}{4 \sqrt{\pi} \sqrt{S}}, 
\qquad
Q(S,U) = -\left(\frac{\partial H}{\partial U}\right)_{S} 
= \frac{\sqrt{S} U}{\sqrt{\pi}}.
\end{equation}
Furthermore, the enthalpy potential in Eq.~\eqref{RN-enthalpy} remains a quasi-homogeneous function, exhibiting the following scaling properties. The condition $\nu_U = 0$ indicates that the electric potential is a truly intensive thermodynamic variable.
\begin{equation}
    \nu_U = 0, \quad \nu_S = 2 \nu_H, \label{escal_enthal}
\end{equation}
and the Euler identity $\nu_S T S - \nu_U UQ = \nu_H H$ is still satisfied. Consequently, the  metric $g^I$, defined in Eqs.~(\ref{g1}), can be written as
\begin{equation}
 g^I = \nu_H H 
\left(
\frac{-1 + U^2}{8 \sqrt{\pi} S^{3/2}} dS^2
-\frac{  U}{ \sqrt{\pi}  \sqrt{S}}  dS  dU
-\frac{\sqrt{S}}{\sqrt{\pi}}
 dU^2
\right). \label{g1_enthalpy}
\end{equation}
Using Eqs.~\eqref{g1_enthalpy} and~\eqref{g2_energy}, we unify the curvature singularities of both metrics, showing that they are described by the same heat capacity
\begin{equation}
   I,II:H_{,SS} H_{,UU}- (H _{,SU})^2=M_{,SS} M _{,QQ}=\frac{T}{C_Q \times \kappa_S}=0.
\end{equation}
 Within the GTD, thermodynamic interactions are encoded in the curvature of the equilibrium manifold \cite{quevedo2007geometrothermodynamics}. The GTD curvature scalar characterizes the nature of the underlying microscopic interactions: a positive value ($\mathcal{R}^{}>0$) corresponds to repulsive interactions, a negative value ($\mathcal{R}^{}<0$) corresponds to attractive interactions, while a vanishing curvature ($\mathcal{R}^{}=0$) indicates the absence of interactions, as in an ideal gas, which is described by a flat thermodynamic geometry \cite{ruppeiner1979thermodynamics}. Therefore, to investigate the nature of the thermodynamic interactions at the horizon, we compute the corresponding Ricci scalars and analyze their behavior. These scalars admit a reduced representation in terms of the dimensionless parameter $x = S/(\pi Q^2)$, which highlights their universal properties, and leads to
\begin{equation}
\mathcal{R}_{RN}^{I}=\frac{16}{\nu_H Q^{2}}\frac{x^{2}}
{(1-x)(x-3)^{2}}, \label{RN_R1}
\end{equation}
\begin{equation}
\mathcal{R}^{II}_{RN}=\frac{8}{\nu_MQ^{2}}\frac{x^{2}\left(x^{2}+x-4\right)}
{\left(x-3\right)^{2}\left(x+1\right)^{3}},\label{RN_R2}
\end{equation}
\begin{equation}
\mathcal{R}_{RN}^{III}
=-\frac{64}{\nu_M Q^2} \frac{x^2(1+x)}{\left(13-14x+5x^2\right)^2}.\label{RN_R3}
\end{equation}
Notice that $\mathcal{R}^{I}$ is computed from the enthalpy potential and subsequently evaluated at $U \to \sqrt{\pi}Q/\sqrt{S}$. From Eqs.~\eqref{RN_R1}--\eqref{RN_R3}, we observe, as expected, that the parameter $Q$ does not affect the thermodynamic properties of the geometry.
\begin{figure}[h]
\centering
\begin{minipage}[t]{0.32\linewidth}
    \centering
    \includegraphics[width=\linewidth,height=8cm,keepaspectratio]
    {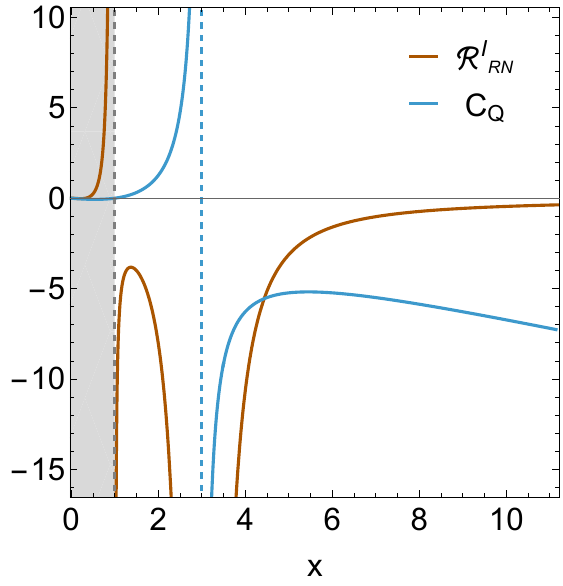}\\
    (a)
\end{minipage}
\hfill
\begin{minipage}[t]{0.33\linewidth}
    \centering
    \includegraphics[width=\linewidth,height=8cm,keepaspectratio]
    {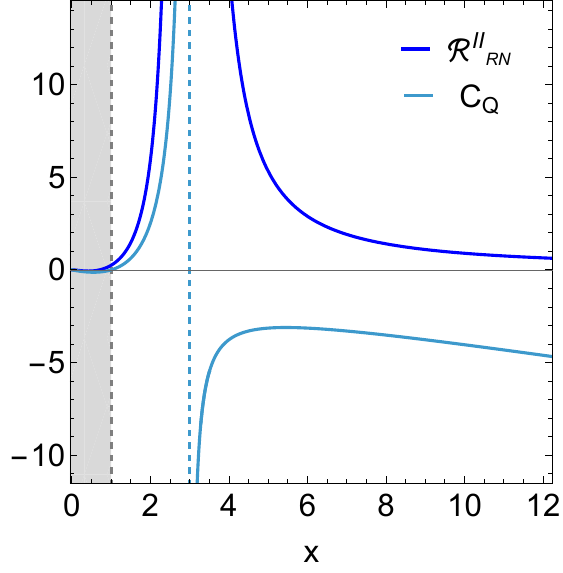}\\
    (b)
\end{minipage}
\hfill
\begin{minipage}[t]{0.32\linewidth}
    \centering
    \includegraphics[width=\linewidth,height=8cm,keepaspectratio]
    {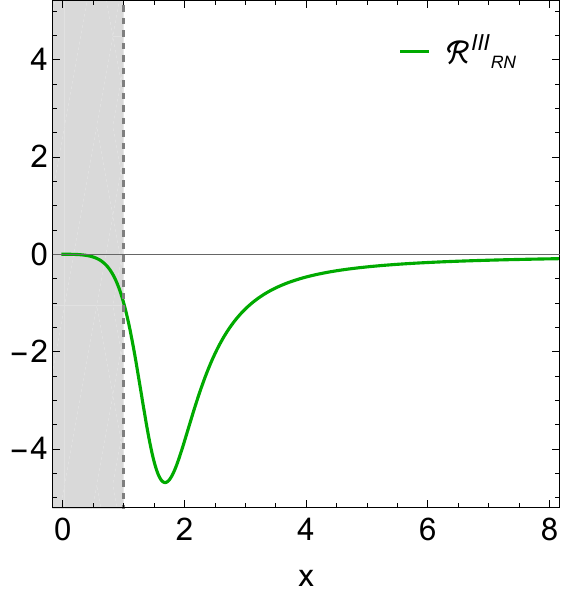}\\
    (c)
\end{minipage}

\caption{Reduced GTD scalars and heat capacity of the RN black hole. 
(a) \( \mathcal{R}^{I}\) computed from the potential $H(S,U)$,
(b) \( \mathcal{R}^{II} \) and (c) \( \mathcal{R}^{III} \) computed from the potential $M(S,Q)$. 
Gray regions indicate non-physical configurations where $T < 0$. 
All plots have been rescaled for clarity.}

\label{RN_scalar}
\end{figure}
We observe that Fig.~\ref{RN_scalar} is consistent with the results summarized in Table~\ref{Table I}. The scalar curvatures $\mathcal{R}^{I}$ and $\mathcal{R}^{II}$ correctly capture the divergence associated with the heat capacity $C_{Q}$. Moreover, $\mathcal{R}^{I}$ predicts a hypothetical phase transition between the extremal black hole configuration ($T = 0$) and a naked singularity. Finally, we find that $\mathcal{R}^{III}$ remains regular throughout the entire parameter space and fails to reproduce the phase structure of the black hole. Consequently, in what follows, we focus exclusively on $\mathcal{R}^{I}$ and $\mathcal{R}^{II}$. In addition, Fig.~\ref{RN_scalarR2_microstructure} shows the curvature scalars as functions of  temperature. We observe that for $Q \neq 0$ the scalar curvature becomes multivalued, indicating the existence of two distinct thermodynamic branches that merge at $T_{\max}$. In the limit $Q \to 0$, the Schwarzschild solution is recovered, resulting in a thermodynamic behavior that mimics a single fluid phase \cite{Ladino:2024ned,ladino2025phase}.
Next, we analyze the behavior of both curvature scalars in the vicinity of $T_{\max}$ in order to extract the predicted power-law scaling of the reduced GTD scalars. This provides a quantitative criterion to characterize thermodynamic singularities in black hole systems. The associated scaling exponent encodes the nature of the thermodynamic instability. In the RN case, the problem is sufficiently simple to allow for an analytical treatment. Using the parametrization of the two thermodynamic branches given in Eq.~\eqref{x_branches}, we obtain 
\begin{align}
\nonumber
\mathcal{R}^{I}_{s}(\chi)
&=
\frac{
2(-1+\cos\chi)
\left[1-2\cos\!\left(\tfrac{\pi+\chi}{3}\right)\right]^4
}{
\left[
-1+2\cos\chi
+4\bigl(-1+\cos\!\left(\tfrac{\pi+\chi}{3}\right)\bigr)
\cos\!\left(\tfrac{\pi+\chi}{3}\right)
\right]^2
\left[
1+2\cos\chi
+12\bigl(-1+\cos\!\left(\tfrac{\pi+\chi}{3}\right)\bigr)
\cos\!\left(\tfrac{\pi+\chi}{3}\right)
\right]
},
\\[2ex]
\mathcal{R}^{I}_{l}(\chi)
&=
\frac{
2(-1+\cos\chi)
}{
\left(1-2\cos\!\left(\tfrac{\chi}{3}\right)\right)^2
\left(1+2\cos\!\left(\tfrac{\chi}{3}\right)\right)^3
},
\\[2ex]\nonumber
\mathcal{R}^{II}_{s}(\chi)
&=
\frac{
4\left[1-2\cos\!\left(\tfrac{\pi+\chi}{3}\right)\right]^4
\sin^2\!\left(\tfrac{\chi}{2}\right) I(\chi)
}{
\left[
-1+2\cos\chi
+4\bigl(-1+\cos\!\left(\tfrac{\pi+\chi}{3}\right)\bigr)
\cos\!\left(\tfrac{\pi+\chi}{3}\right)
\right]^2
\left[
-11+2\cos\chi
+12\cos\!\left(\tfrac{\pi+\chi}{3}\right)
+6\sin\!\left(\tfrac{\pi+4\chi}{6}\right)
\right]^3
},
\\[2ex]\nonumber
\mathcal{R}^{II}_{l}(\chi)
&=
\frac{
2\bigl(
-14
+9\cos\!\left(\tfrac{\chi}{3}\right)
+9\cos\!\left(\tfrac{2\chi}{3}\right)
-4\cos\chi
\bigr)
}{
\left(1-2\cos\!\left(\tfrac{\chi}{3}\right)\right)^2
\left(-5+2\cos\!\left(\tfrac{\chi}{3}\right)\right)^3
};\nonumber
\end{align}
where for simplicity we have defined
\begin{align}
  I(\chi)&\equiv 
-237-14\cos\chi+8\cos(2\chi)
+378\cos\!\left(\tfrac{\pi+\chi}{3}\right)\\&-42\cos\!\left(\tfrac{\pi+4\chi}{3}\right)
+180\sin\!\left(\tfrac{\pi+4\chi}{6}\right)
-30\sin\!\left(\tfrac{\pi+10\chi}{6}\right). \nonumber
\end{align}
Taking a Taylor expansion in the vicinity of the spinodal point $\chi \to \pi$, we obtain the following asymptotic behavior for the parametrized reduced curvature scalars
\begin{align}
\mathcal{R}^{I}(\chi)
&\sim
-\frac{3}{2\epsilon^{2}}
\pm
\frac{7}{4\sqrt{3}\epsilon}
+\mathcal{O}(\epsilon),
\\[2ex]
\mathcal{R}^{II}(\chi)
&\sim
\frac{15}{16\epsilon^{2}}
\pm
\frac{1}{4\sqrt{3}\epsilon}
+\mathcal{O}(\epsilon),
\end{align}
where $\epsilon \equiv \chi - \pi$, and the sign depends on the particular thermodynamic branch under consideration. Additionally, introducing the reduced temperature variable $\Tilde{T}\equiv T/T_{max}$ and $\tau\equiv(1-\Tilde{T})$, the divergence structure expressed in terms of the parameter $\chi$ can be translated into an equivalent temperature scaling. Indeed, it can be shown directly from Eq.~\eqref{y relation} and Eq. \eqref{maxtemRN} that $\tau \sim \frac{\epsilon^{2}}{8} + \mathcal{O}(\epsilon^{4})  .$
 As a consequence, the scaling behavior of the GTD curvature scalars near $T_{max}$ can be equivalently expressed as
\begin{align}
    \mathcal{R^I}(\tau)\sim -\frac{3}{16 \tau}\pm\frac{7}{8\sqrt{6}\tau^{1/2}}+\mathcal{O}(\tau^{1/2}),
 \\[2ex] 
\mathcal{R^{II}}(\tau)\sim \frac{15}{128\tau}\pm\frac{1}{8\sqrt{6}\tau^{1/2}}+\mathcal{O}(\tau^{1/2}).
\end{align}
Therefore, although the two curvature scalars encode different types of 
thermodynamic interactions, as signaled by $\mathcal{R}^{I}<0$ and 
$\mathcal{R}^{II}>0$, and are computed from different thermodynamic potentials, they
both display the same universal power-law behavior in the vicinity of $T_{\max}$.

\begin{equation}
\mathcal{R}_{\text{GTD}} \sim \tau^{-1} .
\end{equation}
Interestingly, the same scaling behavior was obtained numerically for the apparent cosmological horizon in Einstein gravity when quantum corrections are taken into account \cite{romero2026quasi}. This result points to the existence of a universal thermodynamic structure associated with gravitational horizons. Moreover, it stands in clear contrast to the well-known Ruppeiner power law near the phase transition point, $\mathcal{R}_{\text{Ruppeiner}} \sim \tau^{-2}$, as reported, for example, in \cite{wei2019repulsive}.

\subsection{Kerr Microstructure}
Performing the rescaling of the variables $(S,J)$, one readily verifies that 
the fundamental relation $M(S,J)$ in Eq.~(\ref{fundamenta Kerr}) 
is a quasi-homogeneous function of arbitrary degree $\nu_M$, provided that the condition
\begin{equation}
    \nu_J = \nu_S, 
    \qquad 
    \nu_S = 2\nu_M,
    \label{escal Kerr}
\end{equation}
is satisfied. Under this constraint, the corresponding Euler identity is satisfied
\begin{equation}
    \nu_S T S + \nu_J \Omega J = \nu_M M,
\end{equation}
which confirms the quasi-homogeneous character of the Kerr black hole. Following the results of the previous section and Table~\ref{Table I}, in order to correctly describe the singularities of the heat capacity 
$C_J$, we compute the metric $g^{I}$ directly from the enthalpy potential 
$H = M - \Omega J$
\begin{equation}
H(S,\Omega)
=
\frac{\sqrt{S(\pi-\Omega^2 S)}}{2\pi}.
\end{equation}
The enthalpy representation also preserves the quasi-homogeneous 
structure $\nu_S T S + \nu_\Omega \Omega J = \nu_H H$, with weight relations
\begin{equation}
\nu_S = 2\nu_H,
\qquad
\nu_\Omega = -\frac{1}{2}\nu_S.
\end{equation}
Therefore, the quasi-homogeneous GTD line elements, Eqs.~(\ref{g1})--(\ref{g3}) read
\begin{equation}
\begin{aligned}
g^I &= -\frac{\nu_H}{16(\pi - S \Omega^2)}
\left(
\frac{1}{S} dS^2
+\frac{4 S \Omega (3 \pi - 2 S \Omega^2)}{\pi^2} dS d\Omega
+\frac{4 S^2}{\pi} d\Omega^2
\right), \\[6pt]
g^{II} &= \frac{\nu_M }{4 \pi^2 J^2  + S^2}
\Bigg(
-\frac{48  \pi^4 J^4 + 24\pi^2 J^2  S^2 - S^4}{16 \pi  S^3}  dS^2
+\pi S  dJ^2
\Bigg), \\[6pt]
g^{III} &=
-\frac{\nu_M}{16 \pi S^3 (4 J^2 \pi^2 + S^2)^2}
\Bigg(
\left(
192 J^6 \pi^6
+48 J^4 \pi^4 S^2
-28 J^2 \pi^2 S^4
+S^6
\right) dS^2
\\[8pt]
&\qquad
+8 \pi^2 J S (4 J^2 \pi^2 + 3 S^2)
( 4J^2 \pi^2 + S^2) dS dJ -128 J^2 \pi^4 S^4 dJ^2
\Bigg).
\end{aligned}
\end{equation}
Then, according to Eqs. (\ref{sing con1})--(\ref{sing con2}), the relevant singularities conditions are 
\begin{equation}
   I,II:H_{,SS} H_{,\Omega \Omega}- (H _{,S\Omega})^2=M_{,SS} M _{,JJ}=\frac{T}{C_J \times \kappa_S}=0.
\end{equation}
Furthermore, the GTD scalars in terms of the reduced parameter $x = S/2\pi |J|$ are 

\begin{align}
\mathcal{R}^{I}_{\text{Kerr}}
&=
-\frac{3}{\pi\nu_H|J|}
\frac{
{(3+x^4)(1+x^2)}^2
}{
x\left(x^4-6x^2-3\right)^2
},
\label{Kerr_R1}
\\[6pt]
\mathcal{R}^{II}_{\text{Kerr}}
&=
-\frac{1}{4\pi\nu_M|J|}
\frac{
(1+x^2)(3+x^2)(3+6x^2-5x^4)
}{
x{\left(x^4-6x^2-3\right)}^2
},
\label{Kerr_R2}
\\[6pt]
\mathcal{R}^{III}_{\text{Kerr}}
&=
-\frac{6}{\pi\nu_M|J|}
\frac{
x^3(5-32x^2+176x^4){(1+4x^2)}^4
}{
{\left(1+128x^2+736x^4-2048x^6+4352x^8\right)}^2
}.
\label{Kerr_R3}
\end{align}

\begin{figure}[h]
\begin{minipage}[t]{0.32\linewidth}
 \centering
 \hspace{-4cm} \includegraphics[width=1\linewidth]{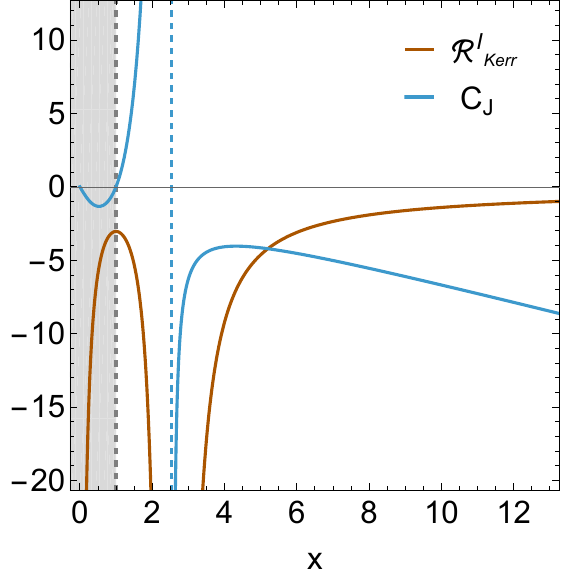}
 \hspace{-5cm}(a)
\end{minipage}%
\hfill%
\begin{minipage}[t]{0.32\linewidth}
 \centering
 \hspace{-4cm} \includegraphics[width=1\linewidth]
 {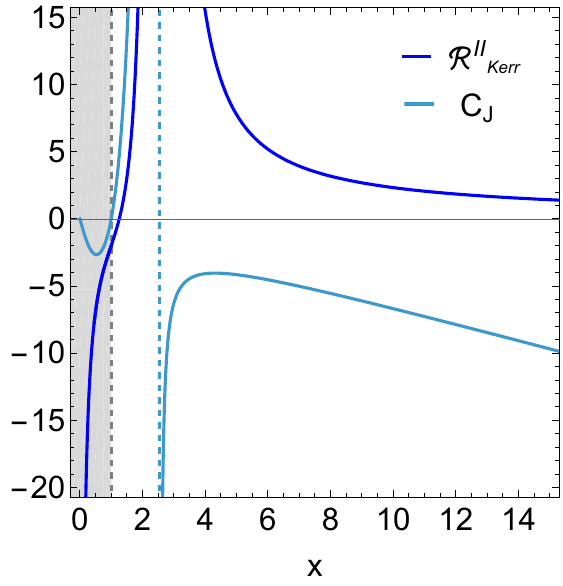}
\hspace{-5cm}(b)
\end{minipage}%
\hfill%
\begin{minipage}[t]{0.32\linewidth}
 \centering
 \hspace{-4cm} \includegraphics[width=1\linewidth]{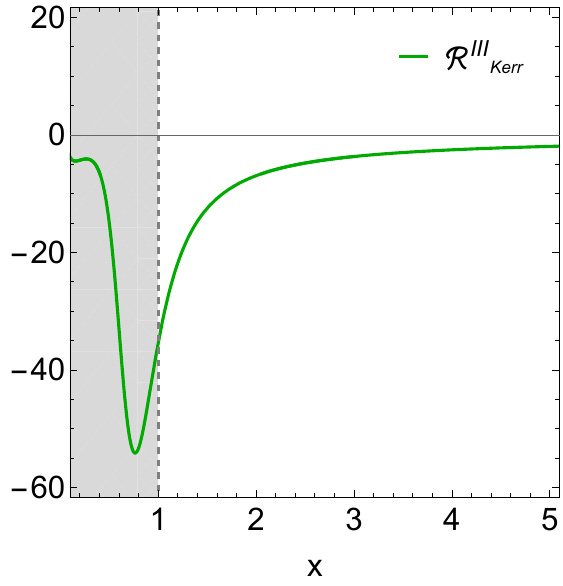}
\hspace{-5cm}(c)
    \end{minipage}%
\hfill%
\caption{Reduced GTD scalars and heat capacity of the Kerr black hole. 
(a) \( \mathcal{R}^{I}\) computed from the  potential $H(S,\Omega)$,
(b) \( \mathcal{R}^{II} \) and (c) \( \mathcal{R}^{III} \) computed from the potential $M(S,J)$. 
Gray regions indicate non-physical configurations where $T < 0$. 
All plots have been rescaled for clarity.}
\label{R Kerr}
\end{figure}
\begin{figure}[h]
\centering
\begin{minipage}[t]{0.34\linewidth}
    \centering
    \includegraphics[width=\linewidth,height=8cm,keepaspectratio]{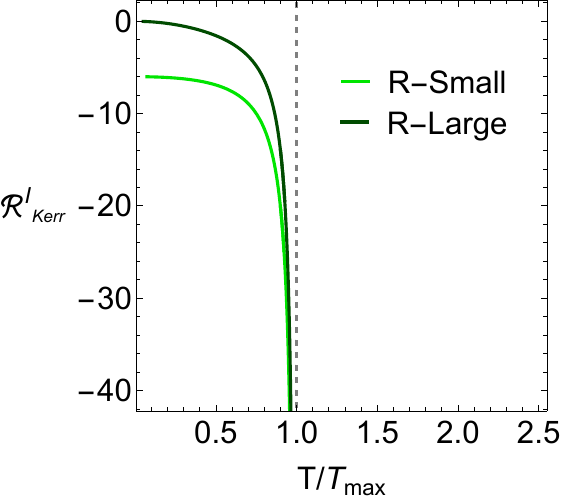}\\
    (a)
\end{minipage}
\hfill
\begin{minipage}[t]{0.32\linewidth}
    \centering
    \includegraphics[width=\linewidth,height=8cm,keepaspectratio]{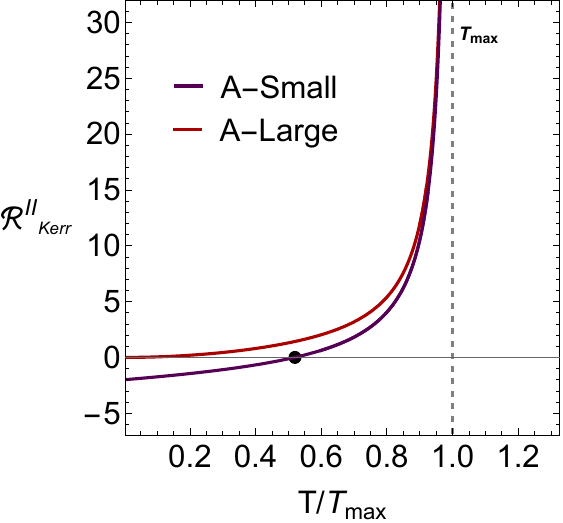}
    (b)
\end{minipage}
\hfill
\begin{minipage}[t]{0.31\linewidth}
    \centering
    \includegraphics[width=\linewidth,height=8cm,keepaspectratio]{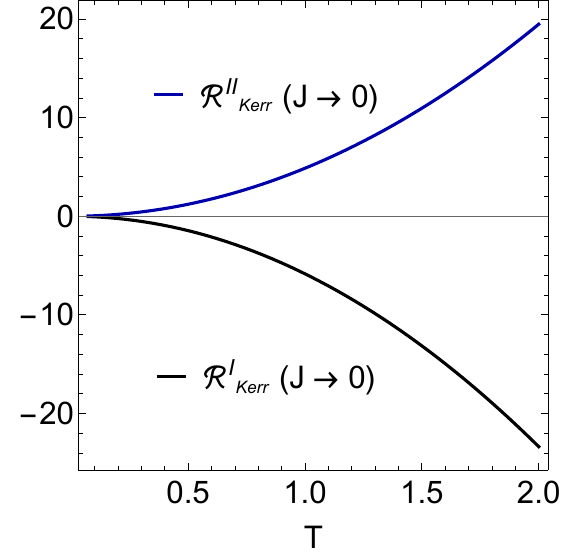}\\
    (c)
\end{minipage}

\caption{GTD scalars of the Kerr black hole as functions of the temperature.
(a) $\mathcal{R}^{I}$, (b) $\mathcal{R}^{II}$, and (c) comparison of both scalars in the Schwarzschild limit $J \to 0$. Here, $R$ and $A$ stand for repulsive and attractive interactions, respectively.}
\label{Kerr scalar_microstructure}
\end{figure}

Fig.~\ref{R Kerr} displays the GTD scalar in the reduced parameter space, while Fig.~\ref{Kerr scalar_microstructure} shows its temperature dependence. As in the RN case, the scalar curvatures reproduce the behavior summarized in Table~\ref{Table I}, capturing the thermodynamic instability associated with $C_J$. Unlike the RN case, $\mathcal{R}^{I}_{\mathrm{Kerr}}$ remains finite in the extremal limit $T \to 0$ (see Figs.~\ref{R Kerr}(a) and~\ref{Kerr scalar_microstructure}(a)). Moreover, unlike in the RN case, $\mathcal{R}^{II}_{\mathrm{Kerr}}$ vanishes within the physical region (Figs.~\ref{R Kerr}(b) and~\ref{Kerr scalar_microstructure}(b)). This zero does not signal criticality, typically associated with singularities, but instead corresponds to a vanishing heat capacity, indicating a change in the local thermodynamic response. In the Schwarzschild limit, Fig.~\ref{Kerr scalar_microstructure}(c) coincides with the RN result in Fig.~\ref{RN_scalarR2_microstructure}(c), where in both cases $\mathcal{R} \sim \pm T^2$ up to a numerical factor, confirming the universality of the Schwarzschild microstructure regardless of whether the limit is approached from rotating or charged configurations.

Finally, we determine the universal power-law behavior of the GTD scalars of the Kerr black hole near the thermodynamic instability. In this case, the thermodynamic branches $x(T)$, implicitly defined  by Eq.~\eqref{kerr branches}, cannot be obtained in  an analytic closed form. Nevertheless, taking advantage of the reduced scalars given in Eq. \eqref{Kerr_R1} -- Eq. \eqref{Kerr_R2}, together with the reduced maximum temperature Eq.~\eqref{maxtemKerr}, we perform a numerical analysis in the vicinity of $T_{\mathrm{max}}$ by fitting the curvature to the scaling ansatz \cite{romero2026quasi}
\begin{equation}
    \mathcal{R}(T) \sim \frac{A}{\left(T - T_{\text{max}}\right)^{\zeta}}.
\end{equation}

\begin{table}[h]
\centering
\begin{tabular}{c|c|c|c|c}
\hline
 & $\mathcal{R}^I$-L 
 & $\mathcal{R}^I$-S
 & $\mathcal{R}^{II}$-L 
 & $\mathcal{R}^{II}$-S \\
\hline
$A$      &-0.0018  &-0.0020  &0.0089  &0.0087  \\
$\zeta$  &1.0036  &0.9960  &0.9996  &1.0013  \\
\hline
\end{tabular}
\caption{Numerical scaling of the reduced GTD scalars $\mathcal{R}^I$ and $\mathcal{R}^{II}$ for the Kerr black hole near $T_{\max}$.}
\label{tab:fit_kerr}
\end{table}
From Table~\ref{tab:fit_kerr}, we observe that the fitted scaling exponent is $\zeta \simeq 1$, independently of the GTD scalar considered and of the thermodynamic branch approaching the instability. This behavior signals a universal linear divergence of the GTD curvature in the vicinity of $T_{\mathrm{max}}$. The fact that $\zeta$ remains invariant under changes in geometric representation and thermodynamic branch further supports the universality of the curvature divergence in the Kerr black hole, mirroring the behavior previously observed in the RN case.

\section{Black Hole Shadows in Entropy Space}\label{sec:shadows}
A key feature in images of supermassive black holes is a bright ring surrounding a dark region known as the shadow. This region is not the event horizon itself, but results from photon trajectories captured by the black hole, creating a brightness depression in the image. Photons with impact parameters slightly above a critical value approach unstable spherical orbits, remain temporarily near the black hole, and eventually escape to infinity, enhancing the observed brightness and forming the photon ring, while those below the critical value fall into the horizon. The boundary of the shadow on the observer’s image plane, often called the critical shadow curve, is determined by the critical impact parameters of unstable photon orbits. This critical shadow is a purely geometric property of the spacetime, depending only on the gravitational theory and the black hole solution parameters, and not on the underlying astrophysical effects \cite{EHT5-1, Perlick_2022}. Therefore, to probe the strong-field gravity regime, we determine below the critical shadow curves of RN and Kerr black holes.\\

Using the fundamental relation $M(S,E_1,\dots)$, photon trajectory quantities can be rewritten in terms of the entropy $S$, so that both thermodynamic and shadow properties depend on the same variables $(S,E_1,\dots)$. This establishes a direct correspondence between a given shadow and quantities such as temperature or thermodynamic curvature. We therefore begin by characterizing the critical shadow curve for each solution in entropy space to investigate its phase structure and microstructure.

\subsection{Reissner-Nordström Shadows}
In static, spherically symmetric spacetimes, the black hole shadow is determined by the photon sphere radius $r_{ps}$, located outside the event horizon $r_+$ and composed of unstable photon orbits. The shadow appears as a dark silhouette formed by gravitational capture and deflection of light near the horizon (as shown in Fig.~\ref{raytrace}), with size characterized by the shadow radius $R_{sh}$, which always exceeds $r_{ps}$. Photon trajectories can be confined to an arbitrary equatorial plane without loss of generality, and from the Euler-Lagrange equations the null geodesic equations governing their propagation are
\begin{align}
&\dot{t}=\frac{\mathbf{E}}{f(r)}, \qquad \dot{r}=\sqrt{\mathbf{E}^2-\frac{\mathbf{L}^2}{r^2} f(r)}, \label{radial} \qquad \dot{\varphi}=\frac{\mathbf{L}}{r^2},
\end{align}
where $\mathbf{E}$ and $\mathbf{L}$ are the specific energy and the specific angular momentum of the photon, and the dotted quantities denote derivatives with respect to an affine parameter. We introduce the well-known impact parameter $b = \mathbf{L}/\mathbf{E}$, and by setting $\dot{r}=0$, we obtain the effective potential from the radial equation \eqref{radial} as
\begin{equation}
V_{\mathrm{eff}}=\frac{1}{b}=\frac{\sqrt{f(r)}}{r} .
\end{equation}
From the potential maximum, the photon sphere radius of the RN black hole is
\begin{equation}
f\left(r_{ps}\right)-\frac{1}{2} r_{ps} f^{\prime}\left(r_{ps}\right)=0 ,\quad \quad r_{ps} = \frac{1}{2}\left(3 M+\sqrt{9 M^2-8 Q^2}\right)
\label{photonshadow1}
\end{equation}
where the prime denotes derivatives with respect to $r$. For an asymptotically flat spacetime, such as the RN metric, the critical impact parameter $b_{cr}$ corresponds to the radius of the photon sphere as seen by an observer at spatial infinity ($r \to +\infty$). Equivalently, it represents the shadow radius measured at large distances \cite{Ladino_2023, Antoniou2023, Vagnozzi_2023, ahmed2026shadow, Karthik2025, Zare_2026, Walia_2024, Lemos_2024, Puli_e_2023, G_mez_2024, Tsukamoto_2024, Khodadi_2022}, which is approximately
\begin{equation}
R_{sh} =  \frac{r_{ps}}{\sqrt{f(r_{ps})}} = 
\frac{\sqrt{2 Q^{2}} \left( 3M + \sqrt{9M^{2} - 8Q^{2}} \right)}
{\sqrt{4Q^{2} - 3M^{2} + M\sqrt{9M^{2} - 8Q^{2}}}},
\label{photonshadow2}
\end{equation}
Therefore, using the fundamental thermodynamic relation $M(S,Q)$ of Eq.~\eqref{fundame1}, the shadow radius and the photon sphere radius in the entropy space can be obtained as
\begin{equation}
R_{sh}(S,Q)=\frac{r_{ps}(S,Q)^2}{\sqrt{Q^2 - 2 M(S,Q) r_{ps}(S,Q) + r_{ps}(S,Q)^2}},
\label{photonshadow3}
\end{equation}
and
\begin{equation}
r_{ps}(S,Q)=\frac{3 \pi Q^2 + 3 S + \sqrt{9 \pi^2 Q^4 - 14 \pi Q^2 S + 9 S^2}}{4 \sqrt{\pi}  \sqrt{S}}.
\label{photonshadow4}
\end{equation}
In the Schwarzschild limit, one recovers $r_{ps}(S,0)=3\sqrt{S}/(2\sqrt{\pi})=3M$ and $R_{sh}(S,0)=3\sqrt{3S}/(2\sqrt{\pi})=3\sqrt{3}M$. The shadow boundary is then reconstructed through the projection onto the celestial coordinates $(\alpha,\beta)$, satisfying $R_{sh}^2=\alpha^2+\beta^2$.

\subsection{Kerr Shadows}
\begin{figure}[h]
 \centering
 \hspace{-0.2cm} \includegraphics[width=1.02\linewidth]{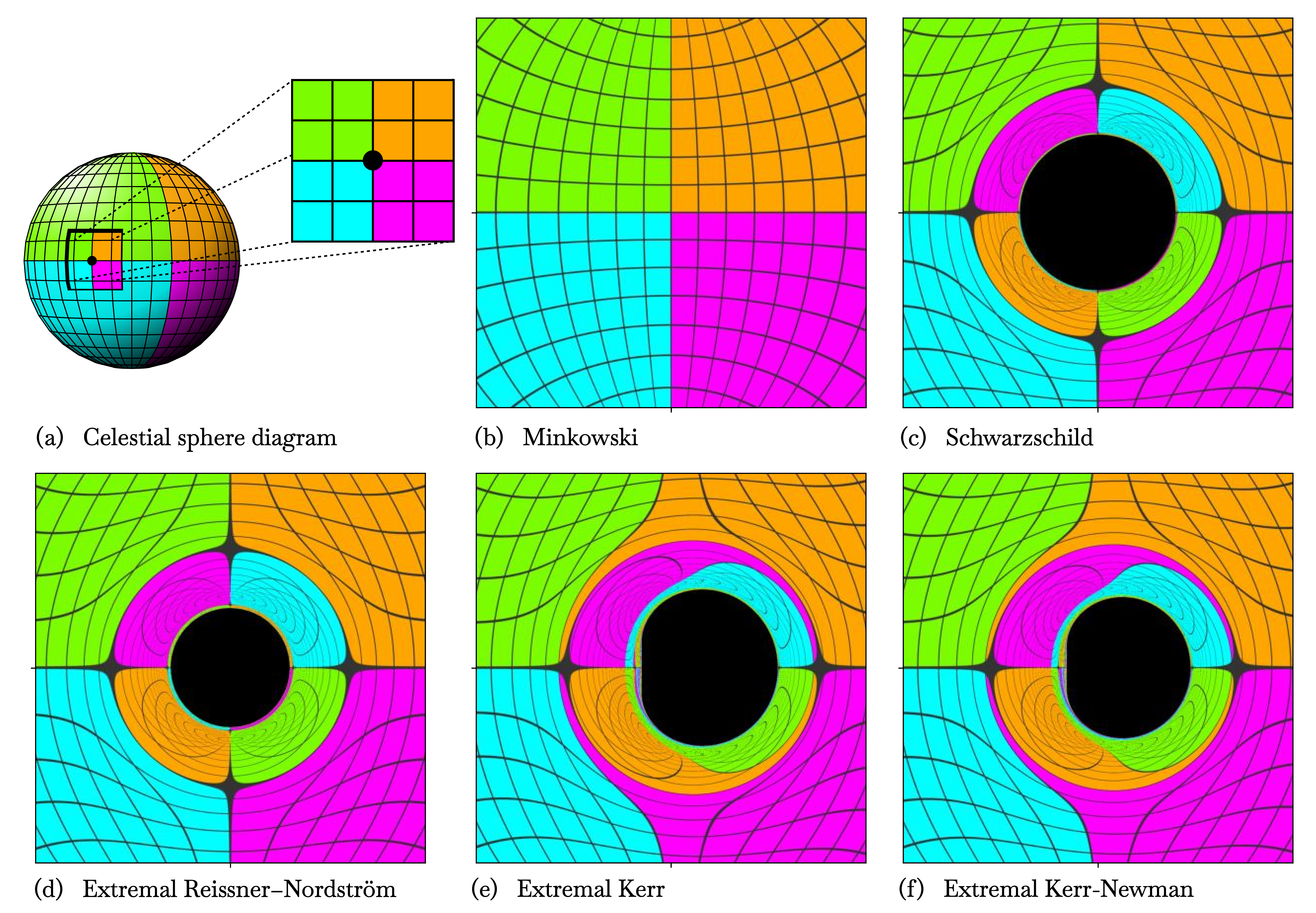}
\caption{ Black hole shadows computed using the ray-tracing tool \texttt{PyHole} \cite{Cunha_2015, Cunha_2016, Cunha_2016_2}, revealing the shadow deformation in rotating spacetimes and the Einstein rings produced by gravitational lensing when the source, black hole, and observer are aligned. In (a), the celestial sphere modeling the ray-tracing background is shown for an equatorial observer ($\theta_0 = \pi/2$) from the interior. The black point indicates the region directly in front of the observer, where the four colored areas intersect. The celestial sphere radius is $R_{\mathrm{EC}} = 30M$, with the observer at $r = 15M$. Light deflection is displayed in (b) Minkowski spacetime and around the black holes: (c) Schwarzschild, (d) extremal RN ($Q=M$), (e) extremal Kerr ($J = M^2$), and (f) extremal Kerr-Newman ($J= \sqrt{3/4}M^2$ and $Q=M/2$).}
\label{raytrace}
\end{figure}

For a static, spherically symmetric black hole, the shadow is perfectly circular. In contrast, rotation breaks this symmetry through frame dragging, yielding a distorted shadow shifted in the direction of rotation (see Fig.~\ref{raytrace}). While in the spherical case unstable photon orbits lie on a single spherical surface, in the rotating case they form a photon region composed of unstable circular orbits at different radii outside the horizon $r_+$. These trajectories can be described using the Hamilton–Jacobi formalism, which provides the null geodesic equations in the Kerr spacetime as \cite{Hioki_2009, Kumar_2020, chandrasekhar, Abdujabbarov_2015}
\begin{align}
\Sigma^2 \dot{t} & =\frac{\left(r^2+a^2\right)^2-a \xi\left(r^2+a^2\right)}{\Delta}- \sin ^2 \theta\left(a^2-\frac{a \xi}{\sin ^2 \theta}\right), \label{motionkerr1}\\
\Sigma^2 \dot{\phi} & =\frac{a\left(r^2+a^2\right)-a^2 \xi}{\Delta}-\frac{a \sin ^2 \theta-\xi}{\sin ^2 \theta},\\
\Sigma^2 \dot{r} & = \pm \sqrt{V}, \quad\quad V(r) \equiv\left[\left(r^2+a^2\right)-a \xi\right]^2-\Delta\left[\left(\xi-a\right)^2+\eta\right], \\
\Sigma^2 \dot{\theta} & = \pm \sqrt{\Theta}, \quad\quad \Theta (\theta)\equiv\left[\eta+\left(\xi-a\right)^2\right]-\frac{\left(a \sin ^2 \theta-\xi\right)^2}{\sin ^2 \theta}, \label{motionkerr4}
\end{align}
where the impact parameters are defined as $\xi \equiv \mathbf{L}/\mathbf{E}$ and $\eta \equiv \mathbf{C}/\mathbf{E}^2$, with $\mathbf{E}$, $\mathbf{L}$, and $\mathbf{C}$ denoting the conserved photon energy, the component of angular momentum along the spin axis and the Carter constant, respectively. The dots indicate derivatives with respect to an affine parameter. The Carter constant reads \cite{Abdujabbarov_2015, Li_2014}
\begin{equation}
\mathbf{C}=(\Sigma \dot{\theta})^2+\cos ^2 \theta\left(\frac{\mathbf{L}^2}{\sin ^2 \theta}-a^2 \mathbf{E}^2\right).
\end{equation}
Photon motion is allowed where $V(r) \geq 0$, leading to scattering, capture by the black hole, or motion along unstable circular orbits. The latter define the photon region around the horizon and shape the black hole shadow. These unstable circular orbits satisfy
\begin{equation}
V(r_{ps}) = 0, \qquad 
\left.\frac{dV(r)}{dr}\right|_{r=r_{ps}} = 0, \qquad 
\left.\frac{d^{2}V(r)}{dr^{2}}\right|_{r=r_{ps}} \ge 0,
\label{condiphoton}
\end{equation}
together with the requirement
\begin{equation} 
\Theta(\theta) \ge 0, \quad \theta \in [0,\pi],
\end{equation}
where $r = r_{ps}$ denotes the radius of each unstable photon orbit. Frame dragging in a Kerr black hole generates nonplanar photon orbits ($\dot{\theta}\neq 0$). Planar circular orbits exist only in the equatorial plane ($\theta=\pi/2$), where the Carter constant vanishes ($\mathbf{C}=0$) \cite{Kumar_2020, Perlick_2022}. These are the prograde (corotating) orbits at smaller radius $r_{ps}^{-}$ and the retrograde (counterrotating) ones at larger radius $r_{ps}^{+}$. Nonequatorial bound orbits are nonplanar, oscillating symmetrically about the equator with $\mathbf{C}>0$, while spherical photon orbits satisfy $r_{ps}^{-} \le r_{ps} \le r_{ps}^{+}$. Thus, $r_{ps}^{\pm}$ bound the photon region that determines the shadow’s critical curve. Setting $a=J/M$ and combining Eqs.~\eqref{condiphoton} with the fundamental relation $M(S,J)$ from Eq.~\eqref{fundamenta Kerr}, we obtain in the equatorial plane \cite{Perlick_2022, Shaikh_2023, Hioki_2009, Abdujabbarov_2015, Li_2014}
\begin{align}
& \xi(S, J) = \frac{M(S, J)^2 r_{ps}^2 \left( r_{ps} - 3 M(S, J) \right) + J^2 \left( M(S, J) + r_{ps} \right)}{J M(S, J) \left( M(S, J) - r_{ps} \right)}, \\
& \eta(S, J) = \frac{M(S, J) r_{ps}^3 \left[ 4 J^2 - M(S, J) r_{ps} \left( r_{ps} - 3 M(S, J) \right)^2 \right]}{J^2 \left( M(S, J) - r_{ps} \right)^2}.
\end{align}
The boundaries of the photon region can also be obtained from the condition $\Theta(\theta)=0$. In particular, unstable equatorial circular orbits are given by the real positive solutions of $\eta(S,J)=0$, from which $r_{ps}^{-}(S,J)$ and $r_{ps}^{+}(S,J)$ are determined by solving
\begin{align}
4 J^2 - M(S, J) \left[ r_{ps}^{\pm} - 3 M(S, J) \right]^2 r_{ps}^{\pm} = 0.
\end{align}
As in ~\cite{LingYi, SINGH2023, Ahmed2025, Zahid2023, Lemos_2024, Ban2025, Shaikh_2023, Kumar_2020}, we employ the celestial coordinates $(\alpha, \beta)$ to represent the black hole shadow, defined as follows
\begin{align}
& \alpha=\lim _{r_0 \rightarrow \infty}\left(-r_0^2 \sin \theta_0 \frac{d \phi}{d r}\right), \qquad \beta=\lim _{r_0 \rightarrow \infty}\left(r_0^2 \frac{d \theta}{d r}\right),
\end{align}
where $r_0$ denotes the distance between the black hole and the observer, and $\theta_0$ represents the inclination angle between the black hole’s rotation axis and the observer’s viewing direction. Calculating $d\phi/dr$ and $d\theta/dr$ from Eqs.~\eqref{motionkerr1}-\eqref{motionkerr4}, and substituting these results into the above expressions in the limit when consider an observer located at spatial infinity $r \to \infty$, we obtain
\begin{align}
\alpha(S, J) &= -\frac{\xi(S, J)}{\sin \theta_0}, \qquad \beta(S, J) = \pm \sqrt{\eta(S, J) + \frac{J^2}{M(S, J)^2} \cos^2 \theta_0 - \xi(S, J)^2 \cot^2 \theta_0}
\end{align}
 Since the Kerr black hole shadow is not perfectly circular but distorted, its size can be characterized through the average shadow radius $R_{sh}$, defined here in a simple manner as
\begin{equation}
R_{sh}(S, J) = \frac{1}{r_{ps}^{+} - r_{ps}^{-}} \int_{r_{ps}^{-}}^{r_{ps}^{+}} \sqrt{(\alpha - \alpha_c)^2 + \beta^2}  dr, \quad \quad
\alpha_c = \frac{1}{r_{ps}^{+} - r_{ps}^{-}} \int_{r_{ps}^{-}}^{r_{ps}^{+}} \alpha  dr
\label{kerrradii}
\end{equation}
where $\alpha_c$ denotes the average horizontal position of the shadow’s center. Fig.~\ref{raytrace} shows, via the ray tracing technique, that increasing the electric charge $Q$ reduces the shadow size, as in the extremal RN and Kerr–Newman cases. In contrast, the effect of angular momentum $J$ is more intricate due to frame dragging. The Kerr shadow preserves its vertical diameter equal to the Schwarzschild value $6\sqrt{3}M$, but contracts horizontally in an asymmetric way, reaching $9M$ in the extremal limit. Motivated by these features, we construct shadow thermodynamic profiles to analyze the thermodynamic properties and microstructure directly from shadow observables.
\begin{table}[h]
\centering
\setlength{\tabcolsep}{5pt}
\renewcommand{\arraystretch}{1.08}
\begin{tabular}{c|c|c|c|c}
\hline
 & $S$ (RN) & $Q$ (RN) & $S$ (Kerr) & $J$ (Kerr) \\
\hline
\begin{tabular}{@{}c@{}}
$S_{\mathrm{ext}}$ \\
($T=0$)
\end{tabular}
& $\pi Q^2$
& $Q=M$
& $2\pi J$
& $J=M^2$ \\
\hline

\begin{tabular}{@{}c@{}}
$S_m$ \\
($C_Q\to\infty$) \\
($C_J\to\infty$)
\end{tabular}
& $3\pi Q^2$
& $Q_m=\sqrt{3}M/2$
& $2\sqrt{3+2\sqrt{3}}\,\pi J$
& $J_m=\sqrt{2\sqrt{3}-3}\,M^2$ \\
\hline

\begin{tabular}{@{}c@{}}
$S_{\mathrm{Nonint}}$ \\
($\mathcal{R}^{II}=0$)
\end{tabular}
& N/A
& N/A
& $2\sqrt{(3+2\sqrt{6})/5}\,\pi J$
& $J_N=\sqrt{\sqrt{6}-3/2}\,M^2$ \\
\hline

\begin{tabular}{@{}c@{}}
$S$ \\
($\mathcal{R}^{I}\to\infty$)
\end{tabular}
& $\pi Q^2$, $3\pi Q^2$
& $Q=M$, $Q=\sqrt{3}M/2$
& $2\sqrt{3+2\sqrt{3}}\,\pi J$
& $J=\sqrt{2\sqrt{3}-3}\,M^2$ \\
\hline

\begin{tabular}{@{}c@{}}
$S$ \\
($\mathcal{R}^{II}\to\infty$)
\end{tabular}
& $3\pi Q^2$
& $Q=\sqrt{3}M/2$
& $2\sqrt{3+2\sqrt{3}}\,\pi J$
& $J=\sqrt{2\sqrt{3}-3}\,M^2$ \\
\hline
\end{tabular}
\caption{
Relevant entropy values for RN and Kerr black holes and their expressions in terms of the mass $M$. Only configurations with $S>0$ and $T\geq0$ are considered.
}
\label{Svalues}
\end{table}
\begin{figure}[h]
\begin{minipage}[t]{0.5\linewidth}
 \centering
 \hspace{-7cm} \includegraphics[width=1\linewidth]{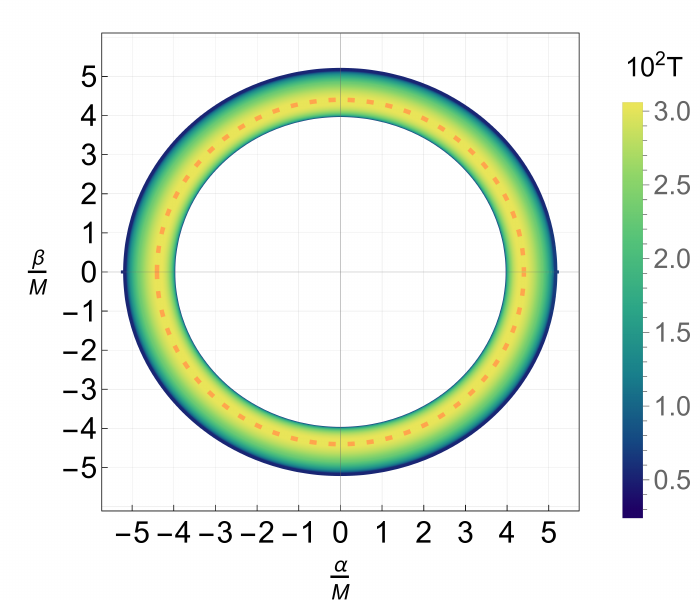}
  \hspace{-8cm} (a)
\end{minipage}%
\hfill%
\begin{minipage}[t]{0.5\linewidth}
 \centering
 \hspace{-7cm} \includegraphics[width=1\linewidth]{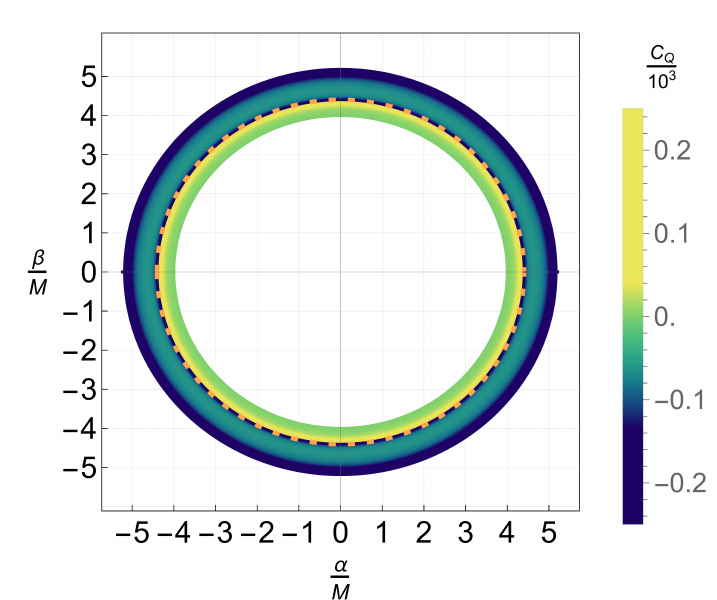}
 \hspace{-8cm} (b)
\end{minipage}%
\hfill \\
\begin{minipage}[t]{0.5\linewidth}
 \centering
 \hspace{-7cm} \includegraphics[width=1\linewidth]{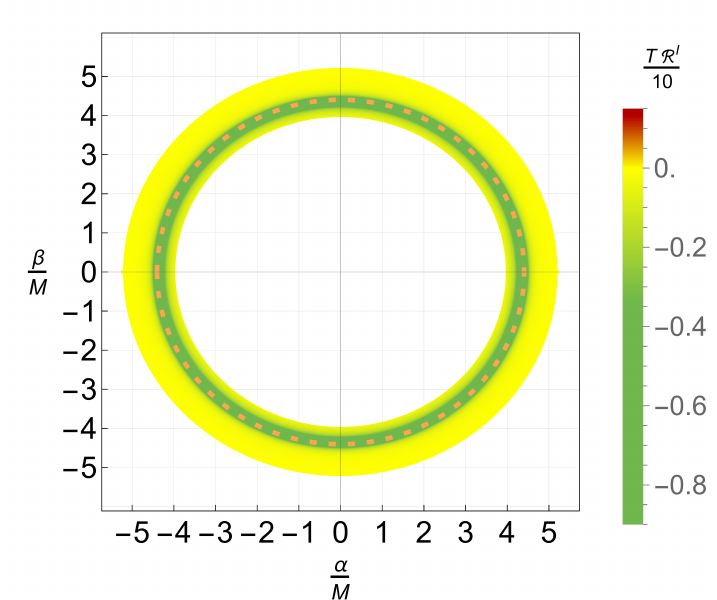}
 \hspace{-8cm}(c)
    \end{minipage}%
\hfill%
\begin{minipage}[t]{0.5\linewidth}
 \centering
 \hspace{-7cm}\includegraphics[width=1\linewidth]{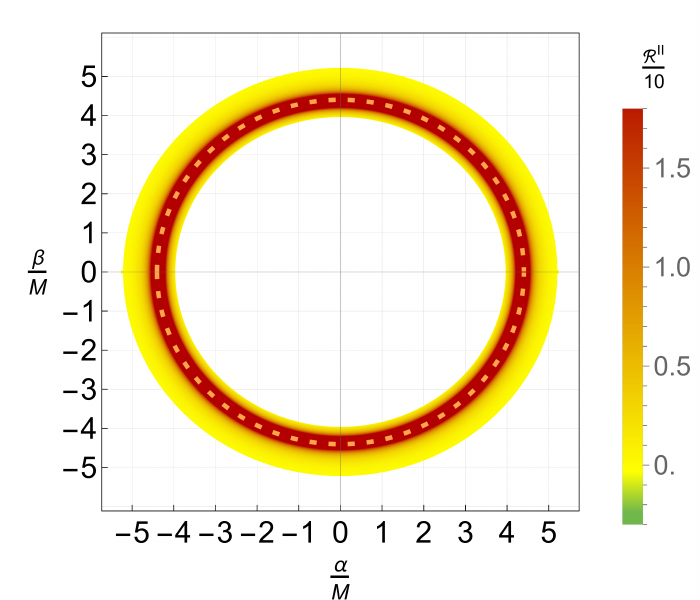}
 \hspace{-8cm}(d)
    \end{minipage}%
\hfill%
\caption{Shadow profiles of the RN black hole: (a) temperature $T$, (b) heat capacity $C_Q$, (c) GTD scalar $\mathcal{R}^{I}$, and (d) GTD scalar $\mathcal{R}^{II}$. The orange dotted line indicates  
(a) $T(S_{m})$,  
(b) $C_Q \to \infty$,  
(c) $\mathcal{R}^{I} \to \infty$, and  
(d) $\mathcal{R}^{II} \to \infty$,  
highlighting their correspondence at the same shadow.  We use $\beta_H =\beta_M = 1$, $Q = M$, and $S \in [S_{ ext}, \infty)$. $T$, $C_Q$, $\mathcal{R}^{I}$, and $\mathcal{R}^{II}$ are expressed in units of $M^{-1}$, $M^{2}$, $M^{-2}$, and $M^{-2}$, respectively. \label{ShadowProfiles1}}

\end{figure}

\subsection{Shadow Thermodynamic Profiles}
\begin{figure}[h]
\begin{minipage}[t]{0.5\linewidth}
 \centering
 \hspace{-7cm} \includegraphics[width=1\linewidth]{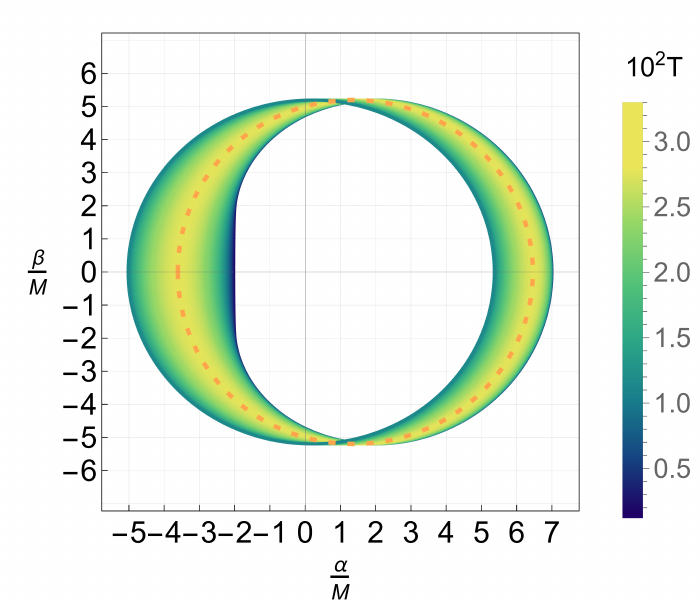}
  \hspace{-8cm} (a)
\end{minipage}%
\hfill%
\begin{minipage}[t]{0.5\linewidth}
 \centering
 \hspace{-7cm} \includegraphics[width=1\linewidth]{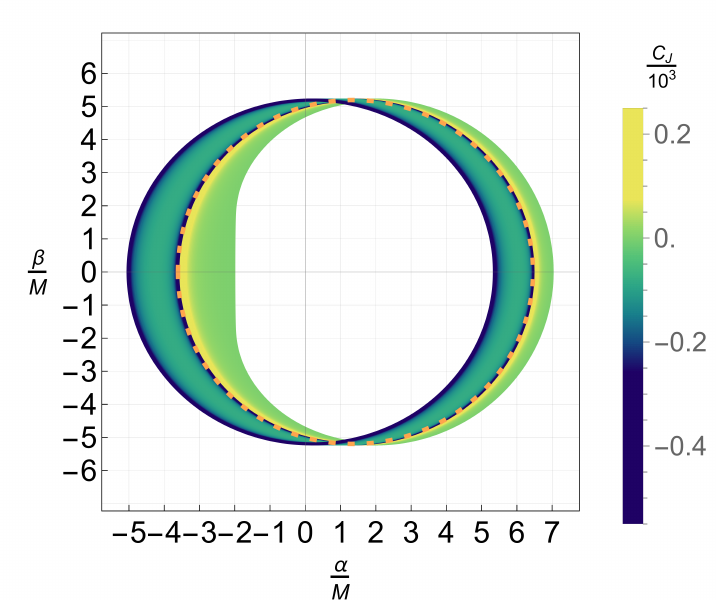}
 \hspace{-8cm} (b)
\end{minipage}%
\hfill \\
\begin{minipage}[t]{0.5\linewidth}
 \centering
 \hspace{-7cm} \includegraphics[width=1\linewidth]{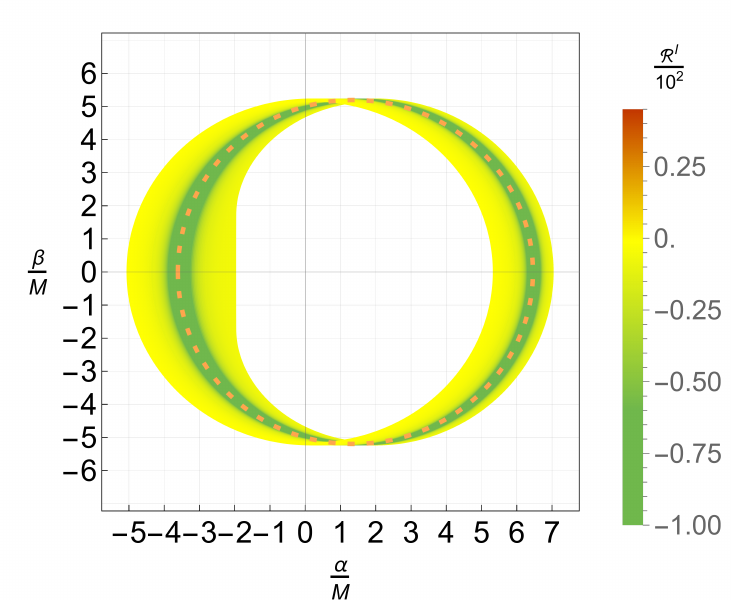}
 \hspace{-8cm}(c)
    \end{minipage}%
\hfill%
\begin{minipage}[t]{0.5\linewidth}
 \centering
 \hspace{-7cm}\includegraphics[width=1\linewidth]{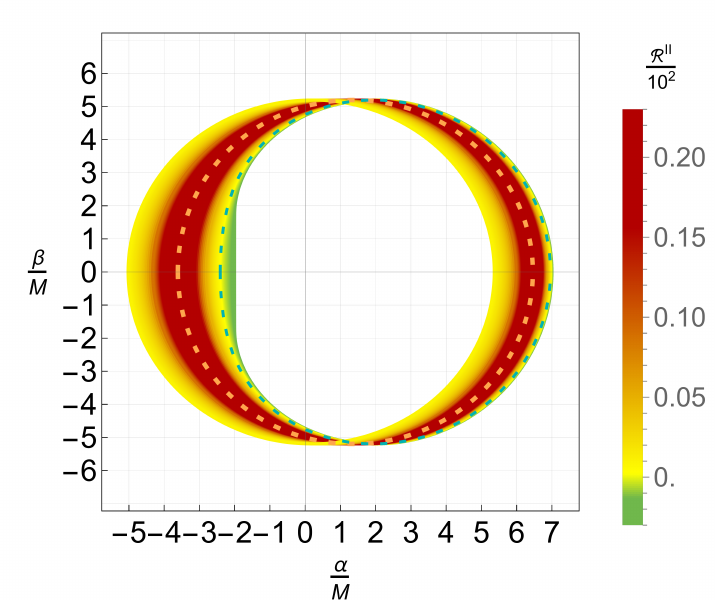}
 \hspace{-8cm}(d)
    \end{minipage}%
\hfill%
\caption{Shadow profiles of the Kerr black hole: (a) temperature $T$, (b) heat capacity $C_J$, (c) GTD scalar $\mathcal{R}^{I}$, and (d) GTD scalar $\mathcal{R}^{II}$. The orange dotted line indicates  
(a) $T(S_{m})$,  
(b) $C_J \to \infty$,  
(c) $\mathcal{R}^{I} \to \infty$, and  
(d) $\mathcal{R}^{II} \to \infty$,  
highlighting their correspondence at the same shadow. The cyan dotted line in (d) represents $\mathcal{R}^{II} = 0$.  We use $\beta_H =\beta_M = 1$, $\theta_0=\pi/2$, $J = M^2$, and $S \in [S_{ ext}, +\infty)$. $T$, $C_J$, $\mathcal{R}^{I}$, and $\mathcal{R}^{II}$ are expressed in units of $M^{-1}$, $M^{2}$, $M^{-2}$, and $M^{-2}$, respectively.
 }

\label{ShadowProfiles2}
\end{figure}

Based on Section~\ref{GTDsection}, we analyze black hole thermodynamics through their shadows by studying the singularities of the GTD curvature scalars $\mathcal{R}^{I}$ and $\mathcal{R}^{II}$, associated with $g^{I}$ and $g^{II}$, which must match those of the heat capacities $C_Q$ or $C_J$. The signs of the scalars further reveal the effective microscopic interactions encoded in the shadow. We implement the GTD framework using $\Phi=H(S,U)$ and $\Phi=M(S,Q)$ for RN, and $\Phi=H(S,\Omega)$ and $\Phi=M(S,J)$ for Kerr, corresponding to $g^{I}$ and $g^{II}$. Table~\ref{Svalues} lists the relevant entropy values for RN and Kerr, expressed in terms of $M$ and restricted to $S>0$ and $T\ge0$. We find exact agreement between the singularities of the heat capacities and those of $\mathcal{R}^{I,II}$, and identify $S_{\text{Noninteractive}}$ and $J_N$, where $\mathcal{R}^{II}=0$, signaling vanishing effective interactions. These features are illustrated in Figs.~\ref{ShadowProfiles1} and~\ref{ShadowProfiles2} through shadow thermodynamic profiles, where critical curves in entropy space are overlaid with a color scale representing the corresponding thermodynamic quantities.
Fig.~\ref{ShadowProfiles1} confirms the correspondence between shadow observables and the thermodynamic phase structure of the RN black hole. For fixed $Q$, the Davies point occurs at $S_m=3\pi Q^2$, separating the SBH phase $S\in[S_{\text{ext}},S_m)$ from the LBH phase $S\in(S_m,\infty)$. Local stability requires $T>0$ and $C_Q>0$, so only the SBH phase admits thermodynamically stable shadow configurations. In particular, the extremal RN shadow ($Q=M$) with $R_{sh}=4M$ lies in the stable SBH branch, whereas in the limit $S\to\infty$ the shadow approaches the Schwarzschild value $R_{sh}=3\sqrt{3}M$, corresponding to the unstable LBH phase with $C_Q<0$. The scalar $\mathcal{R}^{I}$ is rescaled by a factor of $T$ to remove the divergence at $T=0$. Microscopically, $\mathcal{R}^{I}<0$ indicates attractive interactions, while $\mathcal{R}^{II}>0$ signals repulsive ones.

Fig.~\ref{ShadowProfiles2} shows that Kerr black holes display the same correspondence between shadows and thermodynamic structure as observed for RN black holes. For fixed $J$, the Davies point occurs at $S_m=2\sqrt{3+2\sqrt{3}}\pi J$, where the critical shadow curves coincide and thermodynamic quantities become singular. Only the SBH branch is locally stable, including the extremal Kerr case ($J=M^2$), whose highly distorted shadow belongs to this phase. In the limit $S\to\infty$, the shadow approaches the Schwarzschild value $R_{sh}=3\sqrt{3}M$, corresponding to the unstable LBH phase. Microscopically, $\mathcal{R}^{I}<0$ throughout the SBH regime indicates attractive interactions, while $\mathcal{R}^{II}$ changes sign within this phase, revealing repulsive, noninteracting, or attractive effective interactions depending on $S$. The condition $\mathcal{R}^{II}=0$ identifies the noninteracting configuration at $S_{\text{Noninteractive}}$. This result confirms the link between shadow properties and Kerr thermodynamic phase structure.

\section{Thermodynamics and Microstructure from the Shadows of Sagittarius A*}\label{sec:sagi}
The Event Horizon Telescope (EHT) is a global Earth-sized Very-Long Baseline Interferometry array operating at 1.3 mm, which produced the first horizon-scale images of the supermassive black holes in Messier 87 and the Milky Way, known as M87* and Sagittarius A*. In this work, we focus on Sagittarius A*, whose proximity and lower mass relative to M87* allow both a more precise determination of its mass-to-distance ratio and the exploration of a curvature regime intermediate between M87* and stellar-mass black holes \cite{EHT5-1}. The photon ring structure, including the critical shadow curve, provides a stringent test of General Relativity and alternative theories~\cite{Antoniou2023, Filho_2025, LingYi, ESLAM2025, Zare_2026, Walia_2024, Vagnozzi_2023, Cao_2023, Ahmed2025, Walia_2023, Zahid2023, Sahoo, Lemos_2024, Ban2025, Zhao_2024, Puli_e_2023, G_mez_2024, Chenn2024, Tsukamoto_2024, Khodadi_2022, Shaikh_2023,2025ApJ...995..148E,2025PDU....4701734Z,2025EPJC...85.1085W,2025JHEAp..4700367A,2025A&A...693A.265E,2026PDU....5102203G,2025PDU....4701785N,2025PDU....4902017T,2025EPJC...85..878B,2025PhLB..86839812L,2026NuPhB102217212R,2025EPJC...85..973R,2025JCAP...11..069Y,2024PDU....4401501C,2024EPJC...84..136L,2024PDU....4401455W,2024PhRvD.109f4064K,Du_2023, Karthik2025, SINGH2023,Ladino_2023}. Following \cite{Antoniou2023}, we employ two observational bounds on the shadow size of Sagittarius A*: the `EHT-Images'' diameter from \cite{EHT5-1}, obtained with imaging algorithms, and the mG-Rings'' diameter from \cite{EHT5-2}, derived from analytic model fitting.\\

To constrain black hole parameters, we use the Schwarzschild shadow deviation parameter $\delta = R_{sh}/(3\sqrt{3}M)-1$, which measures deviations from the Schwarzschild prediction and tests the consistency of EHT observations with General Relativity and alternative theories~\cite{Antoniou2023, Filho_2025, ESLAM2025, Zare_2026, Vagnozzi_2023, Ahmed2025, Walia_2023, G_mez_2024, Tsukamoto_2024, Shaikh_2023}.The angular shadow size $\theta_g = GM/(Dc^2)$ depends on the mass $M$ and distance $D$, both of which have been precisely determined for Sagittarius A* from stellar-orbit observations by two independent teams, commonly known as the Keck and the Very Large Telescope (VLTI) groups \cite{Tuan, EHT5-1}. The Keck collaboration reported $D = (7953 \pm 50 \pm 32)$ pc and $M = (3.951 \pm 0.047)\times10^{6}M_\odot$\cite{Tuan}, while the VLTI collaboration found $D = (8277 \pm 9 \pm 33)$ pc and $M = (4.297 \pm 0.012 \pm 0.040)\times10^{6}M_\odot$ \cite{refId0}. These measurements imply $\theta_g = (4.92 \pm 0.03 \pm 0.01)\mu\mathrm{as}$ (Keck) and $\theta_g = (5.125 \pm 0.009 \pm 0.020)\mu\mathrm{as}$ (VLTI) \cite{Antoniou2023}.  The deviation parameter $\delta$ then constrains the shadow radius $R_{sh}$. We consider the EHT-Images and mG-Rings observational sets, adopting for each the average between Keck and VLTI results \cite{Antoniou2023, EHT5-1, EHT5-2}. Since for Schwarzschild $R_{sh}=3\sqrt{3}GM/c^2=3\sqrt{3}D\theta_g$, the corresponding $\delta$ values and $1$-$\sigma$ bounds on $R_{sh}$ are available from shadow size estimates \cite{EHT5-1, Vagnozzi_2023, Antoniou2023}. We consider only the $1$-$\sigma$ constraints, as the $2$-$\sigma$ limits are less restrictive. The adopted bounds are summarized as follows
\begin{equation}
\text{EHT images: }
\left\{
\begin{aligned}
\text{VLTI:}    & \quad \delta = -0.08_{-0.09}^{+0.09},
\qquad 4.31M\leq R_{sh} \leq 5.25M ,\\
\text{Keck:}   & \quad \delta = -0.04_{-0.10}^{+0.09},
\qquad 4.47M \leq R_{sh} \leq 5.46M ,\\
\text{\textcolor{fucsia}{Average:}}& \quad \delta = -0.06_{-0.067}^{+0.064},
\qquad \textcolor{fucsia}{4.54M \leq R_{sh} \leq 5.22M }.
\end{aligned}
\right.
\end{equation}
\begin{equation}
\text{mG-Rings: }
\left\{
\begin{aligned}
\text{VLTI:}    & \quad \delta = -0.17_{-0.10}^{+0.11},
\qquad 3.79M \leq R_{sh} \leq 4.88M ,\\
\text{Keck:}   & \quad \delta = -0.13_{-0.11}^{+0.11},
\qquad 3.95M \leq R_{sh} \leq 5.09M ,\\
\text{\textcolor{cyan}{Average:}}& \quad \delta = -0.15_{-0.074}^{+0.078},
\qquad \textcolor{cyan}{4.03M \leq R_{sh} \leq 4.82M }.
\end{aligned}
\right.
\end{equation}
The average values are highlighted in color to distinguish each observational constraint in the diagrams below. The deviation parameter $\delta$ is consistently negative, indicating a shadow smaller than the Schwarzschild case.  In what follows, we construct the corresponding SM diagrams for the RN and Kerr solutions using the observational bounds for Sagittarius A*.

\subsection{Shadow-Microstructure Diagrams}
\begin{figure}[h]
\begin{minipage}[t]{1\linewidth}
 \centering
 \hspace{-17.8cm} \includegraphics[width=0.92\linewidth]{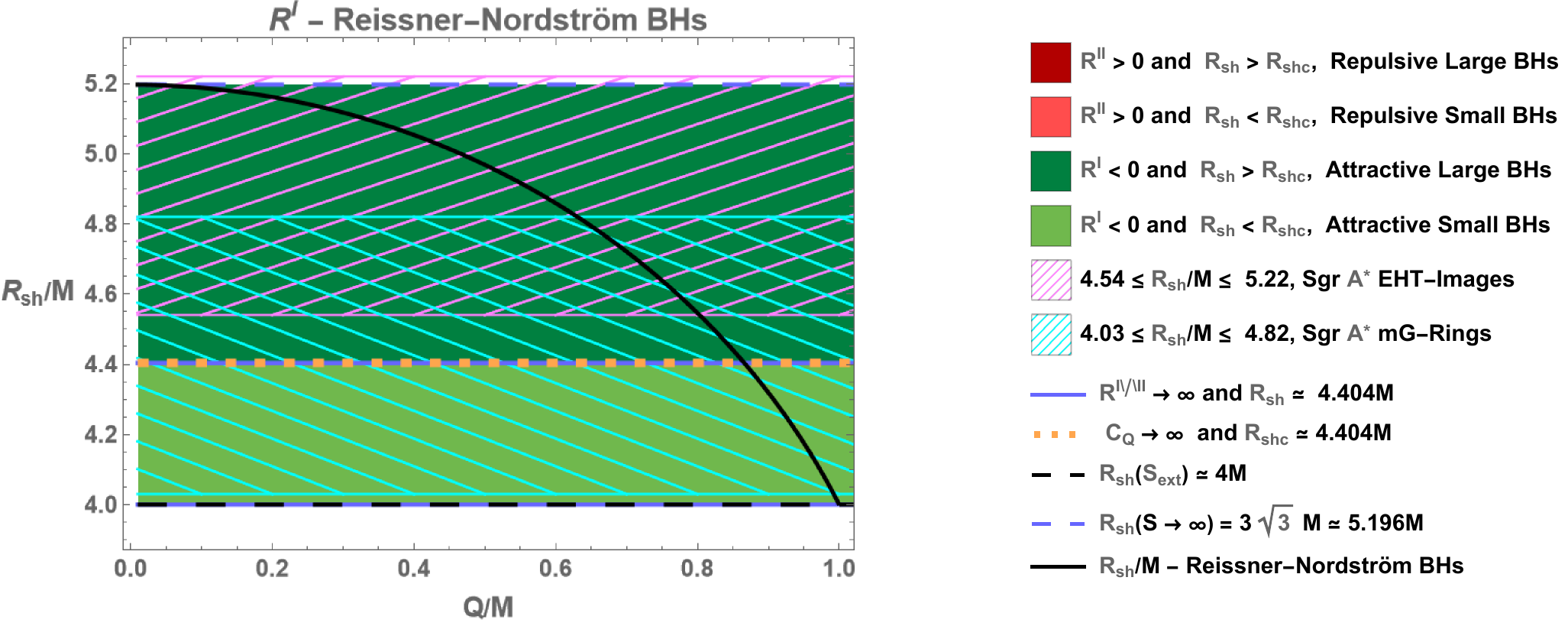}
  \hspace{-16.5cm} (a)
\end{minipage}%
\hfill \\
\begin{minipage}[t]{1\linewidth}
 \centering
\hspace{-8.1cm}\includegraphics[width=1.02\linewidth]{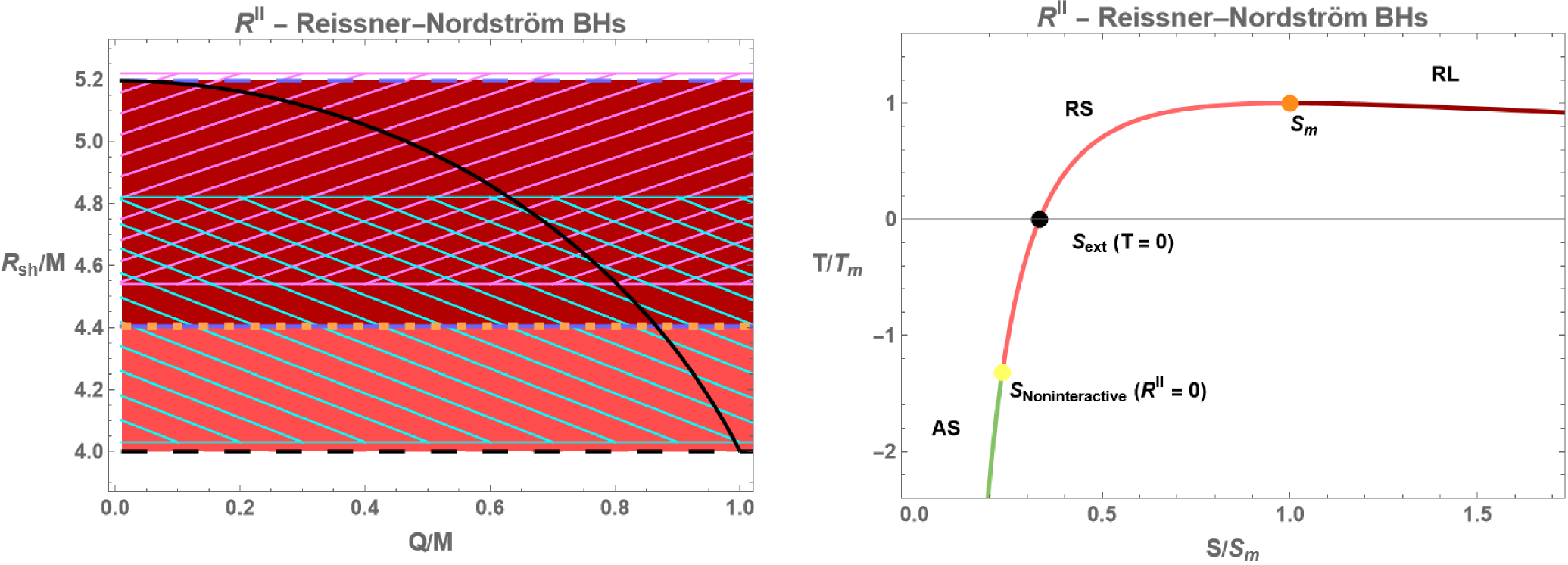}
\hspace{-18cm}(b)\hspace{8.8cm}(c)
    \end{minipage}%
\hfill%
\caption{Shadow–Microstructure diagrams for Sagittarius A* associated with the RN black hole.
(a) Shadow–Microstructure diagram for $\{R_{sh},Q,\mathcal{R}^{I}\}$.
(b) Shadow–Microstructure diagram for $\{R_{sh},Q,\mathcal{R}^{II}\}$.
(c) Temperature $T$ as a function of entropy $S$, highlighting the different interactions associated with $\mathcal{R}^{II}$ and the characteristic points of $S$. The labels AS, RS, and RL denote the Attractive Small, Repulsive Small, and Repulsive Large phases, respectively.}
\label{ShadowDiagramRN}
\end{figure}
\begin{figure}[h]
\begin{minipage}[t]{1\linewidth}
 \centering
 \hspace{-17.2cm} \includegraphics[width=0.95\linewidth]{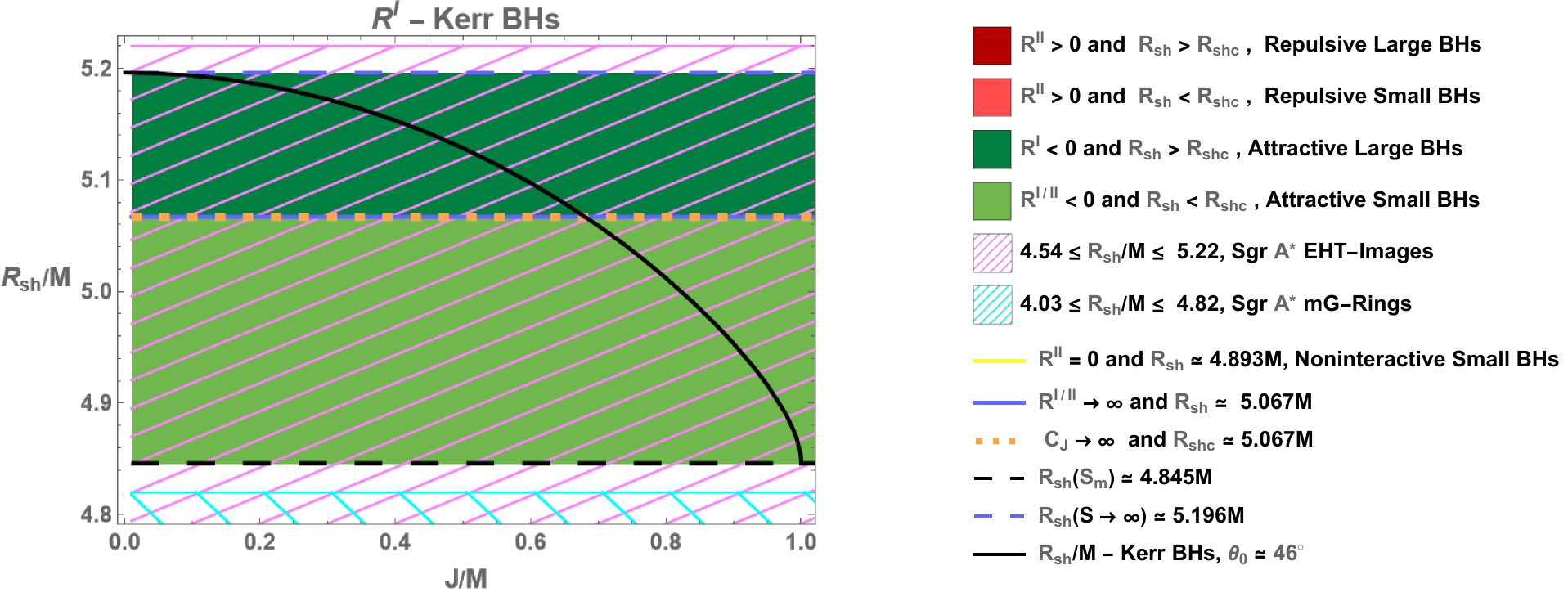}
  \hspace{-16.7cm} (a) 
\end{minipage}%
\vspace{-0.5cm}\\
\hfill \\
\begin{minipage}[t]{1\linewidth}
 \centering
\hspace{-7.8cm} \includegraphics[width=1.01\linewidth]{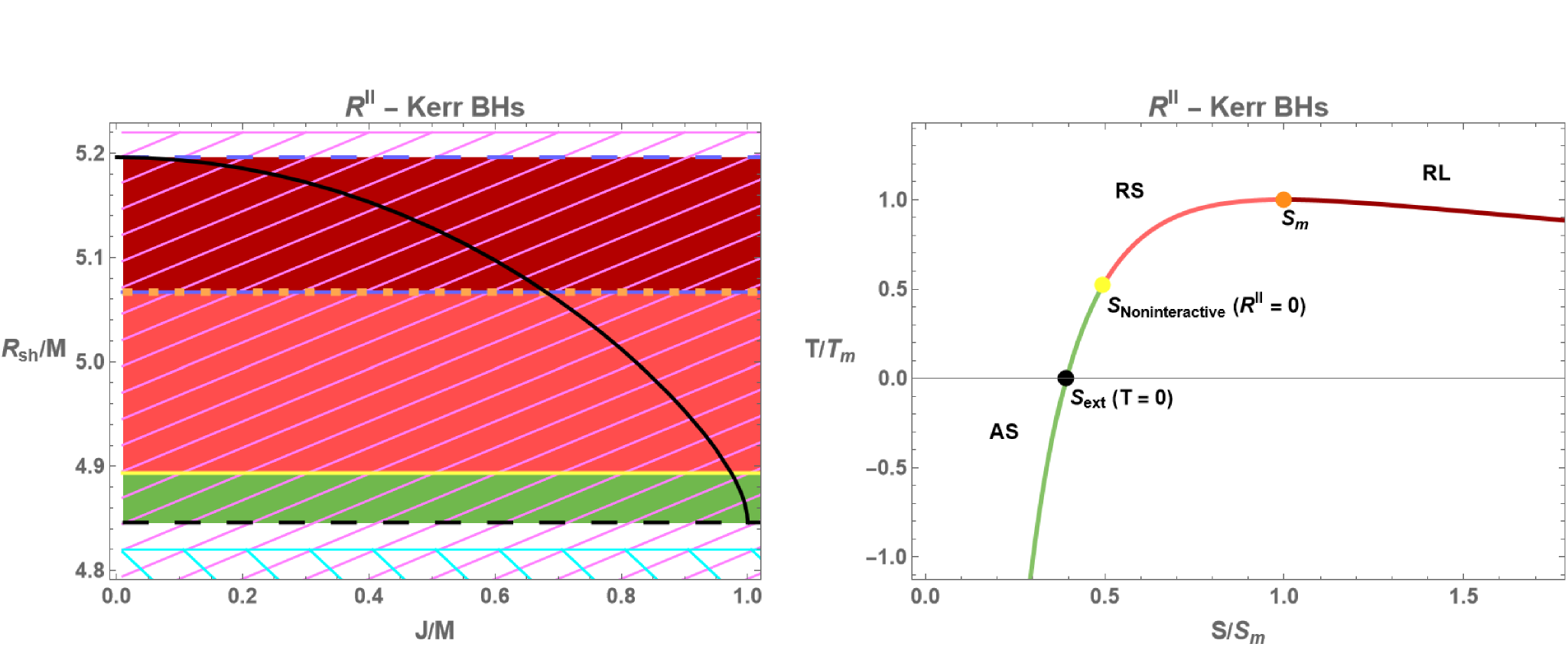}
\hspace{-17.7cm}(b)\hspace{8.8cm}(c)
    \end{minipage}%
\hfill%
\caption{Shadow–Microstructure diagrams for Sagittarius A* associated with the Kerr black hole.
(a) Shadow–Microstructure diagram for $\{R_{sh},J,\mathcal{R}^{I}\}$.
(b) Shadow–Microstructure diagram for $\{R_{sh},J,\mathcal{R}^{II}\}$.
(c) Temperature $T$ as a function of entropy $S$, highlighting the different interactions associated with $\mathcal{R}^{II}$ and the characteristic points of $S$. AS, RS, and RL denote the Attractive Small, Repulsive Small, and Repulsive Large phases, respectively.
 }
\label{ShadowDiagramKerr}
\end{figure}
The connection between black hole shadows and their underlying microscopic thermodynamic interactions can be explored and tested through the Shadow-Microstructure (SM) diagram. We introduce this tool as a region-plot representation that reveals the relationship between a characteristic spacetime parameter $X$, a black hole shadow observable $Y$, and the microscopic interaction measure $Z$. This diagram reveals the possible Microscopic Thermodynamic Phases (MTPs) that a black hole may exhibit through its shadow. By comparing these phases with observational shadow data, one can constrain the parameters of the black hole model and identify the specific MTP associated with an observed black hole. In this work, we define the classification of the MTPs by combining the thermodynamic phase structure (analogous to the small/large transitions in AdS black holes) with the type of microscopic interactions inferred from the sign of the thermodynamic curvature scalars. Accordingly, we distinguish the following phases: Attractive Small (AS), Repulsive Small (RS), Attractive Large (AL), and Repulsive Large (RL), together with the Noninteractive Small (NS) and Noninteractive Large (NL) cases. The SM diagram displays $Y$ as a function of $X$, with color regions on the plane representing the values of $Z$, thereby illustrating the corresponding MTPs consistent with the observational bounds of the shadow. The triplet of quantities $\{X,Y,Z\}$ establishes the framework of the SM diagram. The parameter $X$ characterizes a macroscopic property of the spacetime solution and represents the quantity to be constrained. The observable $Y$ describes the measurable feature of the black hole shadow used for testing, while $Z$ encodes information about the underlying microscopic thermodynamic behavior. To construct the SM diagrams, among the many possible choices for the triplet $\{X,Y,Z\}$, we consider $X$ as either the electric charge $Q$ or the angular momentum $J$, $Y$ as the shadow radius $R_{sh}$, and $Z$ as the sign of the curvature scalars $\mathcal{R}^{I}$ and $\mathcal{R}^{II}$ obtained within the GTD framework. \\

Figure~\ref{ShadowDiagramRN} shows the SM diagrams for Sagittarius A* in the RN spacetime. The shadow radius spans $4M \leq R_{sh} \leq 3\sqrt{3}M$ for $0\le Q\le M$, crossing the colored MTP regions and connecting the Schwarzschild and extremal ($Q=M$) limits. From $\mathcal{R}^{I}$, Sagittarius A* may lie in an AS or AL phase, while $\mathcal{R}^{II}$ indicates RS or RL phases. Imposing the EHT-Images bounds, only the AL phase ($\mathcal{R}^{I}$) and the RL phase ($\mathcal{R}^{II}$) remain viable, whereas all phases are allowed by mG-Rings constraints. In both cases, the singularities of $\mathcal{R}^{I}$ and $\mathcal{R}^{II}$ coincide with $C_Q\to\infty$, intersecting $R_{sh}$ at the critical value
\begin{equation}
R_{shc}=2\sqrt{3+2\sqrt{3}}Q=\sqrt{9+6\sqrt{3}}M\approx4.404M,
\qquad Q_m=\frac{\sqrt{3}}{2}M,
\end{equation}
indicating the critical shadow radius where Sagittarius A* becomes thermodynamically unstable at the Davies point. The associated critical photon sphere radius is $r_{psc}=(1+\sqrt{3})Q=(3+\sqrt{3})M/2\approx2.366M$. Moreover, the noninteracting entropy $S_{\text{Noninteractive}}$ ($\mathcal{R}^{II}=0$) lies in a $T<0$ region, specifically at $S=(\sqrt{41}-5)\pi Q^2/2$ with $T_N\approx-0.040Q^{-1}$. Hence, within the physical domain $T>0$, the NS configuration is not realized and will not be considered in the RN black hole case.\\

Figure~\ref{ShadowDiagramKerr} shows the SM diagrams for Sagittarius A* in the Kerr spacetime. The shadow radius spans $4.845M \leq R_{sh} \leq 3\sqrt{3}M$ for $0\le J\le M$ (with $\theta_0=46^\circ$), crossing the MTP regions. The scalar $\mathcal{R}^{I}$ admits AS and AL phases, whereas $\mathcal{R}^{II}$ allows AS, NS, RS, and RL phases. Unlike the RN case, all phases remain compatible with EHT-image bounds, while mG-ring bounds exclude them. The singularities of $\mathcal{R}^{I}$ and $\mathcal{R}^{II}$ coincide with $C_J\to\infty$, marking the Davies point reached at a critical shadow radius 
\begin{equation} 
R_{shc} \approx 5.928\sqrt{J} \approx 5.067M, \qquad \text{with } J_m = \sqrt{ 2\sqrt{3}-3} M^{2} \approx 0.681M^{2}. \end{equation}
Here, $S_{\text{Noninteractive}}$ lies in the $T>0$ region, so all four phases are admissible. In particular, the extremal Kerr shadow ($J=M^2$ and $R_{sh}\approx4.845M$) and the noninteracting configuration ($J_N\approx0.974M^2$, $R_{sh}\approx4.893M$ and $T_N\approx0.014J^{-1/2}$) both belong to the AS phase.
\subsection{Microscopic Thermodynamic Phases} 
We present in Table~\ref{tab:MTPsSgrA} a summary of our results on the MTPs of Sagittarius~A$^*$ obtained from the SM diagrams constructed for the RN and Kerr black holes, considering both GTD curvature scalars $\mathcal{R}^{I}$ and $\mathcal{R}^{II}$ and imposing the EHT-image and mG-ring observational bounds. These diagrams relate the shadow radius $R_{sh}$ to the relevant black hole parameters, such as the charge $Q$ or the angular momentum $J$, thereby allowing us to identify the MTP regions selected by the observational shadow constraints.
\begin{table}[h]
\renewcommand{\arraystretch}{1}
\setlength{\tabcolsep}{6pt}
\resizebox{\textwidth}{!}{%
\begin{tabular}{c|c|c|c|c|c}
\hline
 \begin{tabular}[c]{@{}c@{}}Black Hole\\ GTD Scalar\end{tabular}                                                                                       &  \begin{tabular}[c]{@{}c@{}} MTPs\\  $[S_{ext}, \infty)$ \end{tabular} 
                                                             & \begin{tabular}[c]{@{}c@{}}MTPs\\ Sgr A*\\ EHT-Images\end{tabular}                     &\begin{tabular}[c]{@{}c@{}}Constraints\\ Sgr A*\\ EHT-Images \end{tabular}                                                                                                                                                                                                          & \begin{tabular}[c]{@{}c@{}}MTPs\\Sgr A* \\ mG-Rings\end{tabular}                 & \begin{tabular}[c]{@{}c@{}}Constraints\\Sgr A* \\ mG-Rings   \end{tabular}                                                                                                                                                                                                                                                                                                                                   \\ \hline
{\begin{tabular}[c]{@{}c@{}}Reissner\\Nordström\\$\mathbf{(\mathcal{R}^{I})}$\end{tabular}}  & \begin{tabular}[c]{@{}c@{}}AL \\ AS \end{tabular}       & {AL}                                                         & {\begin{tabular}[c]{@{}c@{}}\vspace{-0.1in}\\ $12.566 Q^2 \le S \le \infty$ \hspace{0.8cm}    \\ \hspace{0.3cm} $0 \le Q \le 0.803M$ \\ $4.540M \le R_{sh} \le 3\sqrt{3}M$ \\ (AL or RL)\end{tabular}} & {\begin{tabular}[c]{@{}c@{}}AL\\ AS\end{tabular}} & {\begin{tabular}[c]{@{}c@{}} $9.426 Q^2 \le S \le 24.983 Q^2$       \\ \hspace{0.01cm} $0.630M \le Q \le 0.866 M$\\ $4.404M \le R_{sh} \le 4.820M$ \\ (AL or RL) \\  \end{tabular}} \\
                                                                                                     &                                                                                  &                                                              &                                             &                                                                            &                            \\ \cline{1-3} \cline{5-5}
{\begin{tabular}[c]{@{}c@{}}Reissner\\ Nordström\\ $\mathbf{(\mathcal{R}^{II})}$\end{tabular}} & {\begin{tabular}[c]{@{}c@{}}RL\\ RS\end{tabular}}       & {RL}                                                         &                                                                                                                                                                                                                        & {\begin{tabular}[c]{@{}c@{}}RL\\ RS\end{tabular}} &                                                                                                                                                                       {\begin{tabular}[c]{@{}c@{}} \textcolor{red}{$4.049 Q^2 \le S \le 9.426 Q^2$}\hspace{0.1cm}    \\ \hspace{0.01cm}\textcolor{red}{$0.866 M \le Q \le 0.992 M$}\\ \textcolor{red}{$4.030M \le R_{sh} \le 4.404M$} \\\textcolor{red}{ (AS or RS)}\end{tabular}}                                                                                                                                                                           \\
                                                                                                     &                                                                                  &                                                                                  &                                                                 &                                                                            &                                                                                                                               \\ \hline
\begin{tabular}[c]{@{}c@{}}Kerr\\ $\mathbf{(\mathcal{R}^{I})}$\end{tabular}                                    & \begin{tabular}[c]{@{}c@{}}AL \\ AS\end{tabular}                       & \begin{tabular}[c]{@{}c@{}}AL \\ AS\end{tabular}                       & {\begin{tabular}[c]{@{}c@{}}
 \textcolor{black}{$2 \sqrt{3 + 2\sqrt{3}}  \pi J \le S \le \infty$} \\
\textcolor{black}{$0 \le J \le0.681M^2$}\\
 \textcolor{black}{$5.067M \le R_{sh} \le 3\sqrt{3}M$} \\
\textcolor{black}{(AL or RL)}

\end{tabular}}                                                                                                                                                                                         & {\begin{tabular}[c]{@{}c@{}} \\ \\ \\ \\Excluded\end{tabular} }                                                  & {\begin{tabular}[c]{@{}c@{}} \\ \\ \\ \\Excluded\end{tabular} }                                                                                                                                                                                                                                                                                                                          \\ \cline{1-3}
\begin{tabular}[c]{@{}c@{}}Kerr\\ $\mathbf{(\mathcal{R}^{II})}$\end{tabular}                                   & \begin{tabular}[c]{@{}c@{}}RL \\ RS \\ AS \\ NS\end{tabular} & \begin{tabular}[c]{@{}c@{}}RL \\ RS \\ AS \\ NS\end{tabular} & {\begin{tabular}[c]{@{}c@{}}

\hspace{1cm}\\ 
\textcolor{red}{$2 \pi J \le S \le 2 \sqrt{3 + 2\sqrt{3}}  \pi J$} \\ 
\textcolor{red}{$0.681M^2 \le J \le M^2$}\\
\textcolor{red}{$4.845M \le R_{sh} \le 5.067M$} \\
\textcolor{red}{(AS or RS)}\\

\hspace{1cm}\\ 
 \textcolor{red}{$ S_N =2\sqrt{ \frac{1}{5}(3 + 2\sqrt{6})}  \pi J$} \\
\textcolor{red}{$J_N = \sqrt{\sqrt{6}-3/2} M^2$}\\
 \textcolor{red}{$ R_{sh} \approx 4.893M$} \\
\textcolor{red}{(NS)}\\
\end{tabular}}                                                                                                                                                                                                                                                                                                                                              &                                                                            &                                                                                                                                                                                                                                                                                                                                                  \\ \hline
\end{tabular}}
\caption{Summary of the Microscopic Thermodynamic Phases (MTPs) obtained from the shadows of Sagittarius A*.
AS, RS, NS, AL and RL denote the Attractive Small, Repulsive Small, Noninteractive Small, Attractive Large and Repulsive Large phases, respectively. \label{tab:MTPsSgrA}}
\end{table}
Incorporating the EHT-image and mG-ring bounds, the SM diagrams directly constrain the allowed MTPs and parameter ranges. In Table~\ref{tab:MTPsSgrA}, the small phases are highlighted in red as thermodynamically stable.  For RN with $\mathcal{R}^{I}$, EHT bounds select the AL phase within $0\le Q\le0.803M$, while mG-ring bounds restrict the system to $0.630M\le Q\le0.866M$ (AL) and $0.866M\le Q\le0.992M$ (AS), the latter near extremality and stable. For $\mathcal{R}^{II}$, EHT bounds select the RL phase, whereas mG-ring bounds allow both RL and RS phases with analogous $Q$ ranges. For Kerr, mG-ring bounds exclude the configuration for both scalars. Under EHT bounds, $\mathcal{R}^{I}$ allows AL ($0\le J\le0.681M^2$) and AS ($0.681M^2\le J\le M^2$), the latter stable and near extremality. For $\mathcal{R}^{II}$, EHT bounds admit RL, RS, AS, and NS phases within the same $J$ range, with the NS configuration at $R_{sh}\approx4.893M$ and $J_N\approx0.974M^2$, signaling vanishing effective interactions. The Davies point occurs at $R_{shc}\approx4.404M$ with $Q_m\approx0.866M$ (RN) and at $R_{shc}\approx5.067M$ with $J_m\approx0.681M^2$ (Kerr). It lies in regions with attractive interactions according to $\mathcal{R}^{I}$ and repulsive ones according to $\mathcal{R}^{II}$.\\

Our results show that the shadow radius $R_{sh}$ encodes the thermodynamic phase structure and microstructure equivalently to the entropy $S$ description. Applied to Sagittarius A*, $\mathcal{R}^{I}$ favors attractive phases, whereas $\mathcal{R}^{II}$ selects repulsive ones. Larger shadows correlate with AL and RL phases, while smaller ones correspond to AS and RS sectors. Since $\mathcal{R}^{I}$ and $\mathcal{R}^{II}$ lead to distinct microscopic predictions, improved observations could discriminate between GTD metrics, testing thermodynamic formalisms through shadow measurements. Furthermore, thermodynamically stable small phases tend toward near-extremal configurations, so observational bounds favor such regimes when stability is imposed. Thus, shadows may probe not only macroscopic parameters but also microscopic interaction types and stability properties, potentially revealing additional information about the underlying microstructure could be encoded in the behavior of an additional hair in alternative gravity theories beyond the standard macroscopic properties of the black hole solutions of General Relativity~\cite{Antoniou2023,Filho_2025,LingYi,ESLAM2025,Zare_2026,Walia_2024,Vagnozzi_2023,Cao_2023,Ahmed2025,Walia_2023,Zahid2023,Sahoo,Lemos_2024,Ban2025,Zhao_2024,Puli_e_2023,G_mez_2024,Chenn2024,Tsukamoto_2024,Khodadi_2022,Shaikh_2023,2025ApJ...995..148E,2025PDU....4701734Z,2025EPJC...85.1085W,2025JHEAp..4700367A,2025A&A...693A.265E,2026PDU....5102203G,2025PDU....4701785N,2025PDU....4902017T,2025EPJC...85..878B,2025PhLB..86839812L,2026NuPhB102217212R,2025EPJC...85..973R,2025JCAP...11..069Y,2024PDU....4401501C,2024EPJC...84..136L,2024PDU....4401455W,2024PhRvD.109f4064K,Du_2023,Karthik2025,SINGH2023,Ladino_2023}. For Sagittarius A*, the observed bounds correctly capture the spinodal instability at $R_{shc}$. The shadow therefore signals the boundary between stable and unstable regimes. Finally, the NS configuration in Kerr with $\mathcal{R}^{II}$ is particularly intriguing, as it reveals a microscopic state of vanishing effective interactions, signaling the transition between attractive and repulsive regimes and closely resembling ideal gas behavior. This occurs at a characteristic temperature $T_N \approx 0.014J^{-1/2}$, where attractive and repulsive contributions precisely balance each other. By analogy with real gases, where the Boyle temperature corresponds to the vanishing of the second virial coefficient and marks the onset of near-ideal gas behavior \cite{JGPowles_1983,Kovac1998,COCCIA201936}, this point may therefore be interpreted as a Boyle-like temperature $T_{Boyle}$ for Kerr black holes. A deeper understanding of this analogy could be explored in future work through virial expansions applied to thermodynamic systems within GTD.
\section{Conclusions} \label{sec:conclusions}
In this work, we established a consistent correspondence between black hole shadow observables and the thermodynamic phase structure and microstructure of charged and rotating black holes within General Relativity and GTD. We analyzed the thermodynamics and GTD of RN and Kerr black holes. By comparing different thermodynamic potentials, we demonstrated that $g^{I}$ must be constructed from the enthalpy $H(S,I_1)$ and $g^{II}$ from $M(S,E_1)$ to correctly reproduce the phase structure encoded in the heat capacity singularities. Then, expressing the shadow in terms of entropy $S$ and reformulating the thermodynamic analysis through the shadow radius $R_{sh}$, we revealed that the shadow encodes the same phase information conventionally described by the entropy. We constructed and introduced the SM diagrams as a useful tool, showing that they faithfully reproduce the complete thermodynamic phase behavior, including stability and microscopic interaction types, while directly constraining black hole parameters from the bounds on shadow observables.\\
    
For RN and Kerr black holes, the GTD scalars yield distinct microscopic interpretations: $\mathcal{R}^{I}$ selects attractive phases, while $\mathcal{R}^{II}$ favors repulsive ones. Nevertheless, although they predict different types of microscopic interactions, both scalars exhibit the same power-law behavior in the vicinity of the spinodal point, suggesting the existence of a gravitational–thermodynamic universality. Moreover, the radius of the shadow $R_{sh}$ correlates directly with the MTPs, with larger (smaller) shadows corresponding to large (small) phases, providing an observational probe of stability and microscopic interactions. Using EHT-image and mG-ring bounds for Sagittarius A*, we constrain the allowed phases and parameter ranges. For RN, EHT bounds select AL under $\mathcal{R}^{I}$, whereas mG-ring bounds allow AL and a stable near-extremal AS sector; under $\mathcal{R}^{II}$, the allowed regions correspond to RL and RS with analogous charge intervals. For Kerr, mG-ring bounds exclude the solution for both scalars, while EHT bounds allow AL and AS under $\mathcal{R}^{I}$ (the stable AS phase lying near extremality) and RL, RS, AS, and NS under $\mathcal{R}^{II}$. The NS configuration at $R_{sh}\approx 4.893M$ resembles an ideal gas at $T_N$, where attractive and repulsive contributions balance. By analogy to real gases \cite{JGPowles_1983,Kovac1998,COCCIA201936}, this point can be interpreted as an effective Boyle-like temperature $T_{Boyle}$ for Kerr black holes. This analogy could be further explored in future work via virial expansions in GTD. In both spacetimes, the Davies point appears at a critical radius $R_{shc}\approx4.404M$ (RN) and $R_{shc}\approx5.067M$ (Kerr), which we interpret as a spinodal instability rather than a phase transition, occurring in attractive sectors for $\mathcal{R}^{I}$ and repulsive ones for $\mathcal{R}^{II}$. Stable small phases systematically approach extremality, so observational constraints combined with stability naturally favor near-extremal configurations.\\
    
In general, our results show that black hole shadow observations constrain not only macroscopic parameters but also encode information about thermodynamic stability and underlying microscopic interactions. Future high-precision measurements could therefore discriminate not only between black hole solutions but also between different GTD metrics and the thermodynamic approach, testing competing thermodynamic formalisms through astrophysical data. Moreover, this shadow thermodynamic approach can be extended to black holes in alternative theories of gravity~\cite{Antoniou2023,Filho_2025,LingYi,ESLAM2025,Zare_2026,Walia_2024,Vagnozzi_2023,Cao_2023,Ahmed2025,Walia_2023,Zahid2023,Sahoo,Lemos_2024,Ban2025,Zhao_2024,Puli_e_2023,G_mez_2024,Chenn2024,Tsukamoto_2024,Khodadi_2022,Shaikh_2023,2025ApJ...995..148E,2025PDU....4701734Z,2025EPJC...85.1085W,2025JHEAp..4700367A,2025A&A...693A.265E,2026PDU....5102203G,2025PDU....4701785N,2025PDU....4902017T,2025EPJC...85..878B,2025PhLB..86839812L,2026NuPhB102217212R,2025EPJC...85..973R,2025JCAP...11..069Y,2024PDU....4401501C,2024EPJC...84..136L,2024PDU....4401455W,2024PhRvD.109f4064K,Du_2023,Karthik2025,SINGH2023,Ladino_2023}, where extra parameters may leave imprints on both macroscopic properties and the inferred microstructure. Finally, we highlight as a promising future direction the role of forthcoming black hole shadow observational projects, including the next-generation Event Horizon Telescope \cite{EventHorizon3} and the Black Hole Explorer mission \cite{Johnson}, in delivering increasingly precise tests of the predicted thermodynamic behavior.

\let\oldaddcontentsline\addcontentsline
\renewcommand{\addcontentsline}[3]{}

\section*{Acknowledgments}
JML and CRF acknowledge support from Conahcyt-Mexico,  grants No. 4020764 and No. 4003366. This work was supported by UNAM-DGAPA-PAPIIT, grant No. 108225 and Conahcyt, grant No. CBF-2025-I-243.

\renewcommand{\addcontentsline}[3]{\oldaddcontentsline{#1}{#2}{#3}}
\appendix

\let\oldaddcontentsline\addcontentsline
\renewcommand{\addcontentsline}[3]{}
 \setlength{\bibsep}{0pt}
\bibliographystyle{unsrt}
\bibliography{referencias.bib}

\end{document}